\documentclass[acmsmall]{acmart}

\usepackage{booktabs}
\usepackage{csquotes}
\usepackage{pifont}

\newcommand{\ra}[1]{\renewcommand{\arraystretch}{#1}}
\usepackage{array}

\AtBeginDocument{%
  \providecommand\BibTeX{{%
    \normalfont B\kern-0.5em{\scshape i\kern-0.25em b}\kern-0.8em\TeX}}}

\setcopyright{acmcopyright}
\copyrightyear{2018}
\acmYear{2018}
\acmDOI{10.1145/1122445.1122456}

\acmJournal{JACM}
\acmVolume{37}
\acmNumber{4}
\acmArticle{111}
\acmMonth{8}

\begin{document}

\title{CommunityClick: Capturing and Reporting Community Feedback from Town Halls to Improve Inclusivity}

\author{Mahmood Jasim}
\affiliation{%
  \institution{University of Massachusetts Amherst}
  \city{Amherst, MA}
  \country{USA}}
\email{mjasim@cs.umass.edu}

\author{Pooya Khaloo}
\affiliation{%
  \institution{University of Massachusetts Amherst}
  \city{Amherst, MA}
  \country{USA}}
\email{pkhaloo@cs.umass.edu}

\author{Somin Wadhwa}
\affiliation{%
  \institution{University of Massachusetts Amherst}
  \city{Amherst, MA}
  \country{USA}}
\email{sominwadhwa@cs.umass.edu}

\author{Amy X. Zhang}
\affiliation{%
  \institution{University of Washington}
  \city{Seattle}
  \state{Washington}
  \country{USA}}
\email{axz@cs.uw.edu}

\author{Ali Sarvghad}
\affiliation{%
  \institution{University of Massachusetts Amherst}
  \city{Amherst, MA}
  \country{USA}}
\email{asarv@cs.umass.edu}

\author{Narges Mahyar}
\affiliation{%
  \institution{University of Massachusetts Amherst}
  \city{Amherst, MA}
  \country{USA}}
\email{nmahyar@cs.umass.edu}

\renewcommand{\shortauthors}{M. Jasim et al.}

\begin{abstract}
  Local governments still depend on traditional town halls for community consultation, despite problems such as a lack of inclusive participation for attendees and difficulty for civic organizers to capture attendees' feedback in reports. Building on a formative study with 66 town hall attendees and 20 organizers, we designed and developed CommunityClick, a communitysourcing system that captures attendees' feedback in an inclusive manner and enables organizers to author more comprehensive reports. During the meeting, in addition to recording meeting audio to capture vocal attendees' feedback, we modify iClickers to give voice to reticent attendees by allowing them to provide real-time feedback beyond a binary signal. This information then automatically feeds into a meeting transcript augmented with attendees' feedback and organizers' tags. The augmented transcript along with a feedback-weighted summary of the transcript generated from text analysis methods is incorporated into an interactive authoring tool for organizers to write reports. From a field experiment at a town hall meeting, we demonstrate how CommunityClick can improve inclusivity by providing multiple avenues for attendees to share opinions. Additionally, interviews with eight expert organizers demonstrate CommunityClick's utility in creating more comprehensive and accurate reports to inform critical civic decision-making. We discuss the possibility of integrating CommunityClick with town hall meetings in the future as well as expanding to other domains.
\end{abstract}



\maketitle

\section{Introduction}
\label{introduction}
Traditional community consultation methods, such as town halls, public forums, and workshops, are the \textit{modus operandi} for public engagement~\cite{mahyar2019deluge, ehsaei2015successful}. For fair and impartial civic decision-making, the inclusivity of community members' feedback is paramount~\cite{gordon2016civic, mahyar2019deluge, torres2007citizen}. However, traditional methods rarely provide opportunities for inclusive public participation~\cite{Mansbridge2006NormsStudy, Levine2005FutureDeliberation, bryson2013designing}. For instance, reticent meeting attendees struggle to speak up and articulate their viewpoints due to fear of confronting outspoken and dominant individuals~\cite{tracy2007contentious, brabham2009crowdsourcing}. This lack of inclusivity in traditional face-to-face meetings results in an uneven representation of community members and often fails to capture broader perspectives of attendees~\cite{innes2004reframing}. As a result, these methods often fall short in achieving the desired exchange of perspectives between government officials and the community~\cite{salgado2014so, Karpowitz2005DisagreementDeliberation, sanders1997against}. Furthermore, meeting organizers grapple with simultaneously facilitating often contentious discussions and taking meeting notes to capture attendees' broader perspectives~\cite{innes2004reframing, Mansbridge2006NormsStudy}. These bottlenecks further obstruct inclusivity and may lead to biased decisions that can significantly impact people's lives~\cite{Mansbridge2006NormsStudy, mahyar2019deluge}. Advancements in computer-mediated technology can address this predicament by creating a communication channel between these entities.

Bryan~\cite{bryan2010real} and Gastil~\cite{gastil2008political} investigated the state of town halls\footnote{In this work, we use the term \emph{Town Hall} to refer to various community consultation approaches where community members convene to meet and discuss civic issues with government officials or decision-makers. Town Halls can take many forms, but the main goal is to establish a communication channel between the officials and the community, to inform the community about civic issues, and often to receive community's feedback~\cite{bryan2010real}.} and demonstrated a steady decline in civic participation due to the growing disconnect between local government and the community. To reengage disconnected, reticent, or disenfranchised community members, researchers in HCI and digital civics\footnote{Digital Civics is an emerging interdisciplinary area that explores novel ways to utilize technology for promoting broader public engagement and participation in the design and delivery of civic services~\cite{kennethdd, Vlachokyriakos, Olivier}.} have offered novel strategies and technological interventions to increase engagement~\cite{mahyar2019deluge, kennethdd, Olivier, Vlachokyriakos, gordon2016civic}. Researchers in this field have proposed several online technologies that made wider participation possible for community members (e.g.,~\cite{mahyar2018communitycrit, decidim:2019, polis:2017, democracyos:2019}). Despite the introduction of such online platforms, government officials and decision-makers predominantly favor traditional face-to-face meetings to create relationships, foster discourse, and conduct follow-up conversations with community members~\cite{mahyar2019deluge, asad2017creating, corbett2018going} to understand their views and aspirations~\cite{ehsaei2015successful, von2005democratizing, innes1999consensus}. However, employing technology to capture attendees' feedback---in particular, silent attendees' feedback---in face-to-face meetings remains largely unexplored. Commonly, feedback in meetings is gathered using voting or polling attendees~\cite{murphy2009promotion, boulianne2018citizen, lukensmeyer21} or taking notes during the meeting~\cite{lukensmeyer2002taking, manuel2017participatory, bryan2010real}. However, voting often restricts attendees to only agreeing or disagreeing, which often does more harm to the richness of the captured feedback from attendees rather than promoting inclusivity~\cite{mahyar2019deluge, bergstrom2009vote}. To help alleviate this problem, prior work mostly focused on automatic speech recognition~\cite{voicea, tur2010calo} and interactive annotations~\cite{banerjee2006smartnotes, kalnikaite2012markup} to help organizers take notes for creating reports. However, these methods rarely preserve the discussion context or improve the inclusivity of attendees' feedback. 

To better understand the needs of attendees and organizers in community consultations and explore how technology can help to address these issues, we conducted a formative study by attending three town halls in a college town in the United States. We surveyed a total of 66 attendees to inquire about their ability to voice their opinions during town halls. 17\% of the attendees (11 responses) expressed that despite being physically present, they could not voice their opinions. They attributed this to factors such as intimidation from other participants, lack of confidence, and fear of interruption. Moreover, we surveyed 20 organizers to identify what could help them to make better use of the town halls to ensure that the public feedback is prioritized in the reports they authored. We found that the organizers often relied on their memories or the notes taken during the meeting to generate reports. However, they struggled to accurately capture or remember the attendees' feedback and important details after the meeting. In effect, these incomplete memories and meeting notes could potentially result in incomprehensive reports. These findings indicated a requirement for better capturing attendees' feedback and preserving meeting discussion details to support organizers in authoring more comprehensive reports.

Based on our formative study, we designed and developed CommunityClick, a system for town halls that captures more inclusive feedback from attendees during the meeting and enables organizers to author more comprehensive meeting reports. We modified iClickers~\cite{iclicker} to allow attendees to silently and anonymously provide feedback and enable organizers to tag the meeting discussion at any time during the meeting. We chose to use iClickers due to their familiarity and ease of use~\cite{morse2010clicker, herreid2006clicker, whitehead2010usingiClicker} as an audience response system (ARS)~\cite{nickerson1993real}. Furthermore, we augmented the automatically-generated meeting transcript by synchronizing it with attendees' feedback and organizers' tags. We also provided an interactive interface where organizers can explore, analyze, and utilize the augmented meeting transcript in a data-driven way and a novel feedback-weighted summary of the meeting discussion to author meeting reports at their convenience. To evaluate the efficacy of our approach, we conducted a field experiment in the wild, followed by eight semi-structured interviews with experienced organizers. Our results demonstrate the efficacy of CommunityClick to give voice to reticent participants to increase their involvement in town halls, capture attendees' feedback, and enable organizers to compile more inclusive, comprehensive, and accurate meeting reports in a way that lends credibility to the report creation process. 

Our key contributions in this work are as follows: 1) using a communitysourcing technology to enable attendees to share their feedback at any time during the meeting by modifying iClickers as a real-time response mechanism, 2) augmenting meeting transcripts by combining organizers' tags and attendees' feedback to preserve discussion context and feedback from a broader range of attendees, 3) applying a novel feedback-weighted summarization method to generate meeting summaries that prioritize community's feedback, 4) developing an interface for organizers to explore and utilize the augmented meeting transcript to author more comprehensive reports, and 5) insights from a field experiment in the wild that demonstrates how technology can be effective in capturing attendees' voices, authoring more inclusive reports, and future directions to enhance our approach.  

\section{Background}
\label{background}
In this section, we describe the current challenges that inhibit inclusivity in town hall meetings. We also discuss existing technologies designed to promote engagement in various meeting scenarios and to help analyze and utilize meeting data.

\subsection{Current Challenges in Town Halls}

Prior investigations by Bryan~\cite{bryan2010real} and Gastil~\cite{gastil2008political} showed a steady decline in civic participation in town halls due to the growing disconnect between local government and community members and the decline in social capital~\cite{putnam2000bowling, costa2003understanding, rahn1998social}. Despite the introduction of online methods to increase public engagement in the last decade~\cite{coleman2001bowling, mahyar2018communitycrit, decidim:2019, polis:2017, democracyos:2019, kim2016budgetmap}, government officials continue to prefer face-to-face meetings to engage the community in the decision-making process~\cite{mahyar2019deluge, button1999deliberative, ehsaei2015successful}. They believe face-to-face meetings facilitate two-way communications between decision-makers and community members that can foster discourse and help them understand the views and aspirations of the community members~\cite{button1999deliberative, ehsaei2015successful, von2005democratizing, mahyar2019deluge, innes1999consensus}. However, constraints such as fixed physical locations for co-located meetings and scarcity of time and resources, limit the efficacy of face-to-face processes and inhibit the ability of officials to make proper use of town halls~\cite{Levine2005FutureDeliberation, Mansbridge2006NormsStudy, innes2004reframing, seltzer2013citizen, baker2007achieving}. Such constraints might repel or alienate citizens for whom traveling or dedicating time for town halls may not be a viable option from the perspective of physical, economical, or intrinsic interest~\cite{misra2014crowdsourcing, irvin2004citizen, seltzer2013citizen, convertino2015large}. While predictors such as education, income, and intrinsic interest help gauge civic engagement~\cite{davies2014civic}, social dynamics, such as shyness and tendency to avoid confrontation with dominant personalities can also hinder opinion sharing in town halls by favoring privileged individuals who are comfortable or trained to take part in contentious public discussions~\cite{brabham2009crowdsourcing, tracy2007contentious}. Specifically, people with training in analytical and rhetorical reasoning often find themselves at an advantage when discussing complex and critical civic issues~\cite{sanders1997against}. As a result, town halls inadvertently cater to a small number of privileged individuals, and silent participants often become disengaged despite physically attending the meetings~\cite{gordon2011immersive}. Due to the lack of inclusivity, the outcome of such meetings often tends to feel unjust and opaque for the general public~\cite{fung2006varieties, corbette2019trust}. 

\subsection{Technological Interventions to Increase Engagements in Town Halls}
To increase broader civic participation, researchers in HCI have proposed both online~\cite{democracyos:2019, mahyar2018communitycrit, decidim:2019, polis:2017, kim2016budgetmap} and face-to-face~\cite{bergstrom2007conversation, keske2010consulting, lukensmeyer2002taking, taylor2012empowering} technological interventions that use the communitysourcing\footnote{Communitysourcing leverages the specific knowledge of a targeted community. It follows the crowdsourcing model, but takes a more direct approach in harnessing collective ideas, input, and feedback from targeted community members to find solutions to complex problems such as civic decision-making.~\cite{brabham2010crowdsourcing, heimerl2012communitysourcing}} approach. For instance, to increase engagement in town halls, some researchers have experimented with audience response systems (ARS)~\cite{kay2009examining, keske2010consulting, murphy2009promotion, boulianne2018citizen}. Murphy used such systems to promote democracy and community partnerships~\cite{murphy2009promotion}. Similarly, Boulianne et al. deployed clicker devices in contentious public discussions about climate change to gauge public opinions~\cite{boulianne2018citizen}. Bergstrom et al. used a single button device where the attendees anonymously voted (agree/disagree) on issues during the meeting. They showed that back-channel voting helped underrepresented users get more involved in the meeting~\cite{bergstrom2009vote}. The America\textit{Speaks}' public engagement platform, 21 Century Town Meeting\textregistered, also used audience response systems to collect feedback and perform straw polls and votes during town halls~\cite{lukensmeyer21}. However, in these works, the audience response systems were used either for binary voting or polling~\cite{bergstrom2009vote, lukensmeyer21}, or to receive feedback on specific questions that expected attendees' feedback on a Likert scale-like spectrum~\cite{likert1932technique}. These restrictions limit when and how meeting attendees can share their feedback. Audience response systems have seen widespread success as a lightweight tool to engage participants, promote discussions, and invoke critical thinking in the education domain~\cite{morse2010clicker, herreid2006clicker, addison2009usingiclicker, keller2007researchclicker, whitehead2010usingiClicker, mollborn2010meetingclicker}. As such, these devices have the potential to provide a communication channel for silent participants to express their opinions without the obligation to verbalize and risk confrontation. We build upon their success in the education domain by appropriating iClickers for the civic domain. We modify and utilize iClickers for town halls to create a mechanism that supports silent and real-time community feedback. 

HCI researchers have proposed research solutions to increase participation in face-to-face meetings such as design charrettes, or group-meetings in general, by using various tools and methods~\cite{mahyar2016udcospace, hunter2011memtable, bergstrom2007conversation, bergstrom2009vote, Shi2017IdeaWallStimuli}. Some researchers used interactive tabletop and large screen surfaces to engage attendees~\cite{mahyar2016udcospace, hunter2011memtable, Shi2017IdeaWallStimuli}. For example, UD Co-Spaces~\cite{mahyar2016udcospace} used a tabletop-centered multi-display environment for engaging the public in complex urban design processes. Memtable~\cite{hunter2011memtable} used a tabletop display to increase attendees' engagement by integrating annotations using a multi-user text editor on top of multimedia artifacts. IdeaWall visualized thematically grouped discussion contents in real-time on a large screen during the meeting to encourage attendees to discuss such topics~\cite{Shi2017IdeaWallStimuli}. However, large displays along with real-time visualizations might distract attendees from concentrating on meeting discussions~\cite{appleton2005gis}, which might lead to less contribution from the participants irrespective of how vocal they are~\cite{gordon2011immersive}. Furthermore, these innovative approaches might be overwhelming for meeting attendees, especially in the civic domain, due to the heterogeneity of participants with a wide spectrum of expertise and familiarity with technology~\cite{costa2003civic, tracy2007contentious, adams2004publicandemocratic}. It might also be impractical and financially infeasible to use expensive tabletops and large interactive displays for town halls organized in a majority of cities. 

\subsection{Technologies to Analyze and Utilize Meeting Data}
Prior works showed that decision-makers often relied on meeting reports generated by organizers to inform public policies that had potential long-term impacts on the community~\cite{lukensmeyer2002taking}. Commonly, organizers generate these reports based on the outcome of voting or polling attendees~\cite{murphy2009promotion, boulianne2018citizen, lukensmeyer21}, and taking notes during the meeting~\cite{lukensmeyer2002taking, manuel2017participatory, bryan2010real}. They often use manual note-taking techniques such as pen-and-paper or text editors~\cite{di1972listening}. However, simultaneously taking notes from the meetings to capture attendees' feedback and facilitating the meeting to guide and encourage discussions is overwhelming and often lead to losing critical information and ideas~\cite{rowe2005typology, piolat2005cognitive}. Sometimes the meetings are audio-recorded and transcribed for reviewing before writing the report. However, such reviewing processes require significant manpower and labor~\cite{lukensmeyer21, lukensmeyer2002taking}. Furthermore, the audio recording themselves do not capture the feedback of reticent participants who did not speak up. 

In general meeting scenarios, researchers proposed improvements to manual note-taking for capturing meeting discussions~\cite{chiu2001liteminutes, kam2005livenotes, meetingKing, davis1998notepals}. For example, LiteMinutes used a web interface that allowed creating and distributing meeting contents~\cite{chiu2001liteminutes}. MeetingKing provided meeting agenda templates to help organizers create reports~\cite{meetingKing}. Similarly, LiveNotes facilitated note-taking using a shared whiteboard where users could write notes using a digital pen or a keyboard~\cite{kam2005livenotes, bennett1991optical}. Another group of researchers experimented with various tools and techniques to provide support for facilitating online synchronous group discussion. For example, SolutionChat provides moderation support to group discussion facilitators by visualizing group discussion stages and featured opinions from participants and suggesting appropriate moderator responses~\cite{lee2020solutionchat}. Bot in the Bunch takes a different approach and propose a chatbot agent that aims to enhance goal-oriented online group discussions by managing time, encouraging participants to contribute to the discussion, and summarizing the discussion~\cite{kim2020bot}. Similarly, Tilda synthesizes online chat conversation using structured summaries and enable annotation of chat conversation to improve recall~\cite{zhang2018tilda}. 

Closer to our approach, some researchers proposed different techniques for automatically capturing meeting discussions without significant manual intervention~\cite{tur2010calo, voicea, banerjee2006smartnotes, iCompassTech}. For example, CALO and Voicea are automated systems for annotating, transcribing, and analyzing multiparty meetings~\cite{tur2010calo, voicea}. These systems use automatic speech recognition~\cite{rabiner1993fundamentals} to transcribe the meeting audio and extract topics, question-answer pairs, and action items using natural language processing~\cite{hirschberg2015advances}. However, automatic approaches are often error-prone and susceptible to miscategorization of important discussion topics~\cite{mcgregor2017moretomeeting, chuang:2012}. To address this problem, some researchers suggested incorporating human intervention to control and compile reports without completely depending on automatically generated results~\cite{banerjee2006smartnotes, kalnikaite2012markup, iCompassTech}. SmartNotes~\cite{banerjee2006smartnotes} and commercial applications such as ICompassTech~\cite{iCompassTech} use this approach where the users can add and revise notes and topics. 
Although these methods used a combination of automatic and interactive techniques, they enabled the utilization of meeting discussions without capturing sufficient circumstantial input. The automatically generated transcript might record discussions but it may not contain feedback shared by all attendees, especially silent ones. Furthermore, these tools are designed for small-scale group meetings and may not be practical in town halls in the civic domain where attendee numbers and requirements can vary significantly~\cite{bryan2010real, lukensmeyer2002taking}. The heterogeneity of attendees in town halls~\cite{costa2003civic, adams2004publicandemocratic}, the lack of inclusivity in sharing their opinions~\cite{Karpowitz2005DisagreementDeliberation, sanders1997against}, the deficient methods to record meeting discussions~\cite{lukensmeyer21}, and limited financial and design pragmatism in existing technologies~\cite{mahyar2016udcospace, Shi2017IdeaWallStimuli} necessitate a closer investigation and innovation in designing technology to address both meeting attendees' and organizers' challenges regarding inclusivity in town halls.

\section{Formative Study: School Building Town Halls}
\label{formative_study}

To inform our work, we wanted to understand the perspectives and needs of organizers and attendees who participate in community consultations. To this end, we attended several town halls, including multiple public engagement sessions in a college town in the United States (U.S.). The agenda for these town halls included discussions regarding new public school buildings where the town authorities wanted the community's feedback on two new proposals that involved millions of dollars worth of renovation or reconstruction. The town authorities made careful considerations to ensure that the public has access to all the information about the proposals by arranging six community-wide engagement sessions organized by professional facilitators. We joined three out of six of these sessions in a span of three months. We decided to investigate this particular case due to the unique characteristics of the town, where there is relatively high citizen engagement on discussions around education and public schools. We also wanted to investigate how people discuss potentially contentious proposals in the town halls that pitted two contrasting ideas for future public school buildings.

\subsection{Participants and Procedures}
We approached both the meeting attendees, who were community members, and the meeting organizers, who included facilitators and town officials. The town halls began with the organizers presenting their proposal to attendees. Afterward, attendees engaged in discussions and shared their ideas, opinions, and concerns about the proposals. The facilitators were responsible for guiding the discussion and collecting the attendees' feedback.They played a neutral role in the discussion and did not influence the attendees' opinions in any way. 

To understand the attendees' perspectives, we conducted surveys after each town hall. The attendees were all residents of the town and attended the meeting of their own volition. We surveyed a total of 66 attendees. We refer to the attendees from our formative study as \textbf{FA} and organizers from our formative study as \textbf{FO}. In the survey, we asked them open-ended questions regarding their motivation to join the town hall meetings, their experiences in these meetings, and their thoughts around what makes these face-to-face meetings successful. We also asked about their ability to voice their opinions in town halls and about their familiarity with audience response technologies. Furthermore, to gain an understanding of what type of semantic tags they wanted to provide during a town hall, we compiled a list of tags based on prior work on meeting discussions or public engagements that focused on characterizing effective participation during synchronous communication between organizers and meeting attendees in both online and face-to-face settings~\cite{zhang2018tilda, nathan2012incase, derek2010collaborative, zhang2017characterizing, im2018deliberation, murray2006incorporating}. The list of tags is presented in Fig.~\ref{fig:survey}(B). We provided this list to the attendees and asked them to rate which of these tags would help them to express their thoughts in town halls. They rated each tag on a 5-point Likert scale~\cite{likert1932technique}. 

During these town halls, we made contact with the organizers and explained our project goals. From these initial contacts, we used the snowball method~\cite{goodman1961snowball} to reach other organizers working across the U.S. to learn more about their practices regarding town hall meeting organization, facilitation, and report creation. We conducted surveys with a total of 20 organizers with an average experience of 10.5 years (min 1 year, max 35 years) in conducting town halls across the U.S. Our survey participants consisted of town administrators, town clerks, senior planners, directors, professional facilitators, and chairs of different town committees. We asked them open-ended questions around their meeting data recording practices including what data do they prioritize when recording, the challenges they face while organizing the meeting simultaneously and recording meeting notes, and their post-meeting data analysis and meeting report generation processes. We also asked about their choice of tools and technologies for recording meeting notes and for creating reports, and what they think are the elements that constitute a good report. To identify representative tags that could help organizers track and categorize meeting discussion, we compiled another list of tags based on prior works~\cite{zhang2018tilda, nathan2012incase, derek2010collaborative, im2018deliberation, murray2006incorporating}. We used a different list from the one we provided to attendees (Fig.~\ref{fig:survey}(A)) because prior work suggest that organizers and attendees need different tags to categorize meeting discussion based on their perspectives. We asked organizers to rate the tags in the order of importance on a five-point Likert scale~\cite{likert1932technique}. The survey questions and list of tags are provided as supplementary materials. 

\subsection{Findings}
\begin{figure}
\includegraphics[width=1\textwidth]{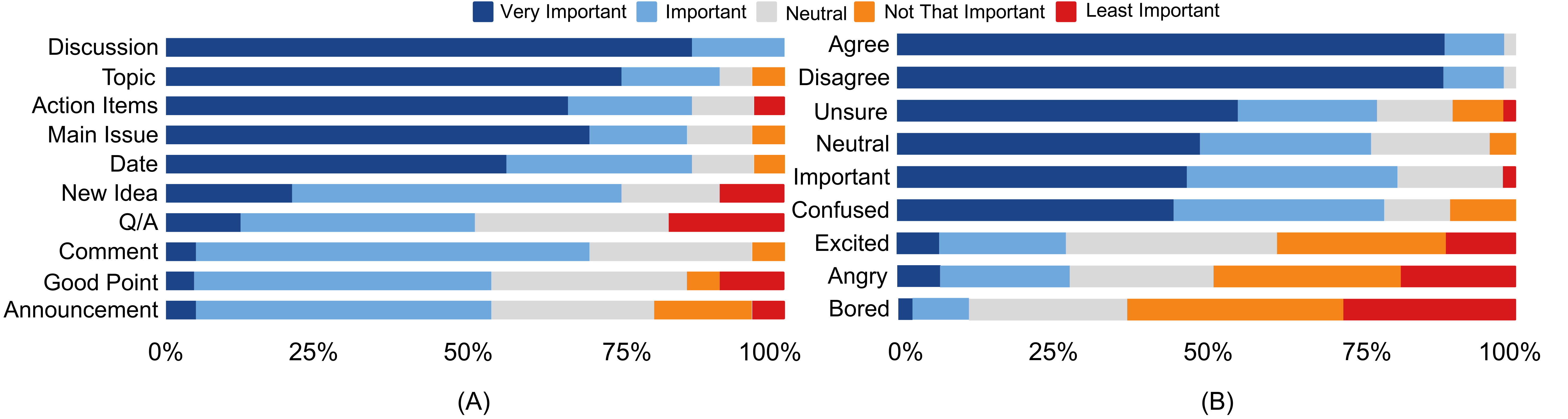}
\caption{This figure presents attendees' and organizers' perceived importance of two sets of tags that we compiled based on prior research: (A) shows the ratings of 20 organizers on a list of 10 tags for organizers; (B) shows the ratings of 66 Community members on a list of 9 tags for attendees' feedback. }
\label{fig:survey}
\end{figure}

Here, we report the findings from our surveys with both attendees and organizers. The 66 attendees we surveyed were highly motivated to attend town halls and the majority of them (64\%, 42 responses) considered attending such meetings to provide their feedback on civic issues as their \emph{civic duty}. Most of the attendees (88\%, 58 responses) attended two or more town halls every year. Regarding their familiarity with technology, every meeting attendee (100\%, 66 responses) mentioned having a computer, smartphone, and internet access, but 61\% of them (40 responses) never used an audience response system before. We also found that 17\% (11 responses) of meeting attendees felt they were not able to share their feedback during these meetings, and 23\% (15 responses) were not satisfied with the way town halls were organized to discuss critical issues. It was surprising for us to find that 17\% of people from a homogeneous, relatively wealthy, and educated community in a college town believed that they could not voice their opinions during town halls. Despite their unfamiliarity with audience response systems, the majority of the meeting attendees (87\%, 57 responses) mentioned that they were willing to use such devices in town halls to share their feedback.

In response to the question regarding what makes face-to-face town hall meeting successful, a group of attendees mentioned that the success of town hall meetings hinges upon the attendees' ability to openly communicate their opinions and hear others' opinions. One attendee (FA-17) mentioned, \blockquote{\emph{Town halls need to provide opportunities for all people to be heard, having diversity of voices and opinions, and be present in the discussion.}} Another attendee (FA-36) mentioned, \blockquote{\emph{Being face-to-face means asking questions, listening to others' questions and ideas, and to be able to see their emotions, nuances and expressions.}} Several attendees' mentioned facing challenges around sharing their opinions due to being dominated in the discussion. One such attendees' mentioned (FA-23), \blockquote{\emph{Town halls give us the chance to talk things out, but often this doesn't happen and people get shut down.}} Another attendee (FA-11) mentioned, \blockquote{\emph{You need to have ground rules so that we stay on track and no one dominates the discussion.}} Some attendees considered that facilitators play an important role in the success of town halls and skilled and well-equipped facilitators can make a difference. One attendee (FA-57) emphasized the importance of skilled facilitator saying, \blockquote{\emph{Professional facilitators understand the context of our community, and the history of tension regarding the issues. They move it along and listen with open minds.}} Another attendee (FA-47) mentioned, \blockquote{\emph{Skilled facilitators make sure all voices are heard, organize the discussion around goals and keep everyone focused on tasks.}}

When rating the tags for sharing opinions during meetings (Fig.~\ref{fig:survey} (A)), the majority (90\%) of the attendees considered \emph{Agree} and \emph{Disagree} to be the most important tags. 75\% or more attendees thought \emph{Unsure}, \emph{Important}, \emph{Confused}, and \emph{Neutral} to be important.

The majority of the organizers (17 responses) mentioned that often manpower and budget constraints forced them to forego the appointment of designated note-takers and thus, they must shuffle between organizing and note-taking during the meetings. Some organizers also mentioned how context-switching between organizing the meeting and taking notes often led to missing critical evidence or information (8 responses). They employed a variety of methods for recording meeting data depending on their convenience and their operational abilities and experiences to utilize such methods, including pen-and-paper (5 responses), text editors (6 responses), audio or video recorders (3 responses), or a combination of these methods. To generate reports from the notes taken during the meetings, organizers usually used text editors (17 responses) with preferences towards Microsoft Word and Notepad (12 responses). 

However, when asked about the time required to compile a report from meeting notes, their responses varied from 15 minutes to a few days. The variation, in part, can be attributed to the amount of notes and the format in which notes were captured. All organizers (20 responses) mentioned that the report generation process involves some form of summarization process of meeting records to retain the most important discussion components. One organizer (FO-7) explained, \blockquote{\emph{If I'm responsible for the report, I listen to the recording, review the notes, and translate them into coherent accounts of meeting attendees, topics, discussion, and decisions/outcomes.}}. Another organizer (FO-1) mentioned, \blockquote{\emph{I start with the agenda, review the audio, then edit and summarize the meeting notes into report content.}} 

Organizers also described high-level properties of what would constitute a \textit{good report}. One organizer (FO-1) mentioned  that, good meeting reports should be \blockquote{\emph{accurate, comprehensive, clear, and concise}}. Another organizer (FO-4) emphasized that several components constitue a good report including, \blockquote{\textit{[the] main agenda items, relevant comments, consensus, action plans, and feedback}}. One organizer (FO-17) thought meeting reports should be \blockquote{\textit{balanced, fair, and pertinent}} towards multiple perspectives, however, it is often challenging because they only \blockquote{listen to a few, while others stay silent}. 

When rating tags that would help them to take better notes to create good reports (Fig.~\ref{fig:survey} (B)), the meeting organizers unanimously (100\%) endorsed the tag \emph{Decision}. The tags \emph{Topic}, \emph{Action Items}, \emph{Main Issue}, and \emph{Date} were preferred by more than 70\% organizers. However, all other tags except for \emph{Q\&A} were favored by more than 50\% of organizers. These survey responses suggest that preferences for tagging meeting discussion varies among organizers, and different sets of tags might be required to be useful for generating reports from different meetings with diverse agendas. 

\subsection{Design Goals}

Based on prior work and our formative study, we identified four design goals to guide our system design that address the requirements and challenges of both meeting attendees and organizers. First, we found that some attendees lacked a way to respond to ongoing discussions. Furthermore, many organizers struggled to keep track of the discussion and take notes simultaneously. Thus, we needed to provide communication channel between attendees and organizers for sharing opinions and capturing feedback (G1). Second, the organizers often refer to several sources of meeting data including meeting notes and audio/video recording to compile meeting reports. Hence, the meeting discussion audio, organizers' tags, and attendees' feedback should be captured and combined together to provide organizers with a holistic view of the meeting data (G2). Third, the organizers perform some form of summarization to generate meeting reports. However, many organizers also struggled to account for attendees' feedback while summarizing meeting data. This challenge motivated a third goal to introduce summarization techniques that incorporate attendees' feedback to help organizers to get the gist of meeting discussions (G3). Finally, organizers needed to examine the meeting-generated data to capture more inclusive attendees' feedback so that they could write more comprehensive reports. To that end, our final design goal was to provide exploration and report authoring functionalities to help them investigate the meeting data and identify evidence to generate reports that included and reflected attendees' feedback (G4).

\begin{figure}
\includegraphics[width=1\textwidth]{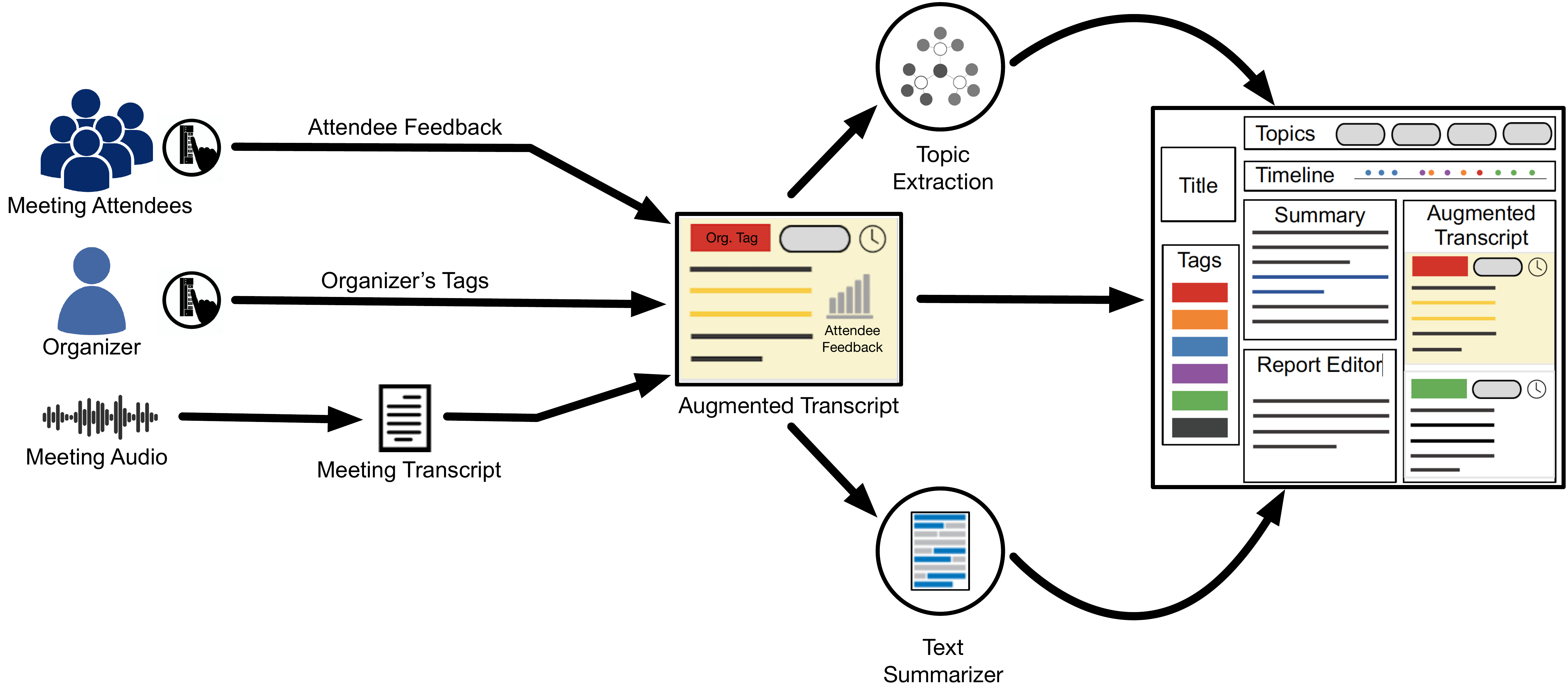}
\caption{A snapshot of CommunityClick's workflow. During the meeting, attendees and organizers can use iClickers to share feedback and tag the meeting respectively. The meeting is also audio-recorded for transcription. The audio recordings are transcribed automatically and then augmented with the organizer's tags and attendees' feedback. Furthermore, we generated the feedback-weighted discussion summary and extracted the most relevant topics. The interactive interface enables the exploration and utilization of augmented meeting discussions, which is available online for organizers to examine and author meeting reports.}
\label{fig:cc_block}
\end{figure}

\section{CommunityClick}
\label{sysdesign}

Guided by the design goals, we designed and developed CommunityClick, a system where we modified iClickers as a real-time response mechanism and augmented meeting transcripts to capture more inclusive feedback. We also introduced a novel feedback-weighted summarization to prioritize attendees' feedback and enabled exploration and utilization of the augmented transcript through an interactive interface for organizers to author meeting reports. Here, we provide a scenario where CommunityClick can be employed (Fig.~\ref{fig:cc_block}), followed by the system description.

\subsection{User Scenario}
Michelle is an organizer who has been appointed by the local government officials to organize an important town hall. Given the importance of the meeting, she decides to deploy CommunityClick to focus on facilitating the meeting while using iClickers to capture the community's feedback. 

Adam is a community member who is attending the town hall. He cares about the community and wants to share his opinions on the agenda. He prefers to avoid confrontations, especially in town halls, as he is worried about speaking up and running into arguments. In the meeting, he is given an iClicker and instructions for using it to share his opinions using five options. Adam engages in discussion with other attendees in the meeting but whenever he feels hesitant to speak up, he uses the iClicker to provide feedback.

A week later, Michelle finally gets around to writing the report of the town hall. By now, she has forgotten a significant portion of the meeting discussion. She logs in to CommunityClick and selects the town hall. She uses the timeline and feedback-weighted summary to get an overview of the meeting discussion and jog her memory by exploring the meeting discussion. She uses her own tags, timeline, and the interactive summary to investigate the augmented meeting transcript that contained attendees' feedback alongside the discussion segments. Finally, she authors the report by importing information into the text editor from the transcript. 

\begin{figure}[h]
\centering
\includegraphics[width=0.5\textwidth]{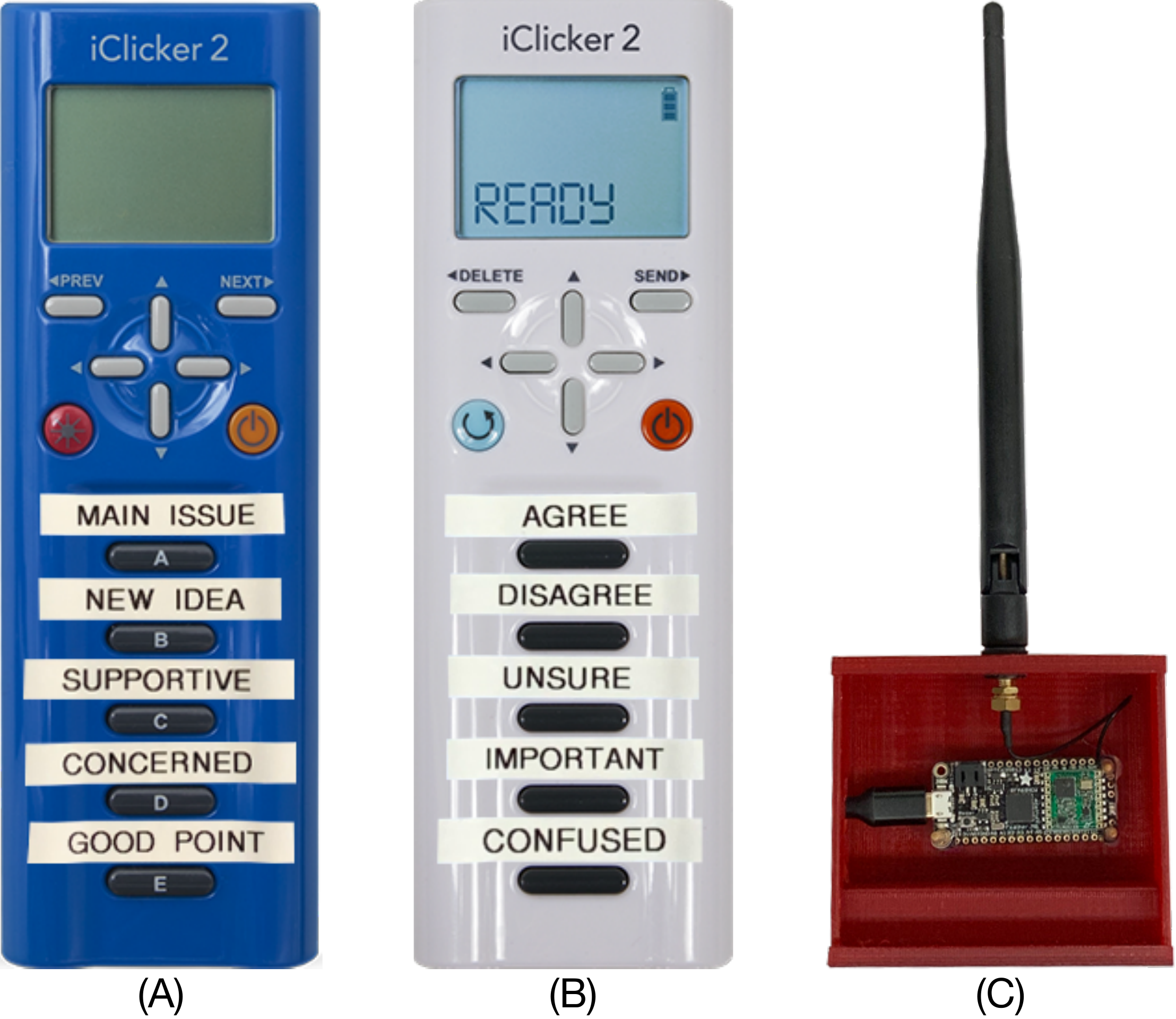}
\caption{The apparatus used to capture organizers' tags and attendees' feedback. (A) The iClicker for organizers to tag the meeting. (B) The iClicker for attendees to add their feedback. We used different sets of tags for organizers and attendees based on our formative study. Each iClicker was labeled with the respective set of tags to reduce the cognitive load of mapping options to the iClicker buttons. (C) The iClicker recorder. We used an Adafruit Feather M0 with the 900 MHz RFM69W/RFM69HW transceiver to capture iClicker clicks with timestamps in real-time to synchronize tags and feedback with meeting audio.}
\label{fig:apa}
\end{figure}

\subsection{System Description}

In the following, we describe how we addressed the design goals by using iClickers, augmenting meeting transcripts, performing text analysis, and developing an interactive interface. 

\subsubsection{Modifying iClickers to enable attendees and organizers to provide real-time responses (G1)}
We used iClickers for both organizers and attendees to enable them to respond to meeting discussions any time during the meeting without the need to manually take notes or speak up to share opinions. iClicker is a communication device that uses radio frequency to allow users to anonymously respond using its five buttons (Fig.~\ref{fig:apa}). Despite the widespread usage of smartphones, we chose iClickers as an audience response system due to recent statistics that show 20\% of U.S. citizens do not yet have access to smartphones~\cite{pewresearch}. Furthermore, they are often a major cause of distraction and hindrance to participation in meetings~\cite{bajko2016prevalence, kushlev2016silence}. There are also technical overheads involved including installation of application, maintenance, and issues regarding version compatibility based on operating systems which might disengage the participants. In contrast, iClickers have proven to be successful in town halls due to its familiarity and affordance of anonymity in sharing opinions in town halls and to receive attendees' feedback on specific questions~\cite{bergstrom2009vote, murphy2009promotion, boulianne2018citizen}. The anonymous use of iClickers could ensure that silent participants can also share their opinions to the organizer about the ongoing discussion without engaging with a potentially heated debate. Moreover, we modified the iClickers to go beyond previous approaches by allowing meeting organizers and attendees to respond to the ongoing discussion using all five different options instead of only binary agree or disagree. The tag options are customizable and depending on their meeting agendas, the organizers can set up an appropriate list of tags and feedback before the meeting. We used different types of iClickers for organizers and attendees. The organizers used instructor iClickers and attendees used regular ones (Fig.~\ref{fig:apa}). It helped us to effectively separate organizers' and attendees' responses. To reduce the cognitive load of iClicker users to map and remember the options, we labeled each iClicker with organizers' tags and attendees' feedback. 

\subsubsection{Combining automatically generated transcripts with tags and feedback to capture more inclusive feedback (G2)}
We recorded and synchronized three different sets of data generated simultaneously in the meeting---the discussion audio, the organizers' tags, and the attendees' feedback via iClickers. We recorded the meeting audio using a regular and commonly used omnidirectional microphone. To remove the noise from the audio recording, we used an open-source freeware named Audacity\textregistered. However, capturing organizers' tags and attendees' feedback from iClickers was non-trivial due to the limitations in hardware access and the API provided by the iClicker manufacturer. The original software and hardware in factory settings did not provide timestamped data on each click. As a result, we customized the hardware and API to record organizers' tags and attendees' feedback (Fig.~\ref{fig:apa}). We used an Adafruit Feather M0 with the 900 MHz RFM69W/RFM69HW transceiver to collect an iClicker's clicks and timestamps that were transmitted through radio frequency on the same bandwidth. This allowed us to accurately and precisely capture and synchronize iClicker interactions to match the time of discussion. 

\begin{figure}
\centering
\includegraphics[width=1\textwidth]{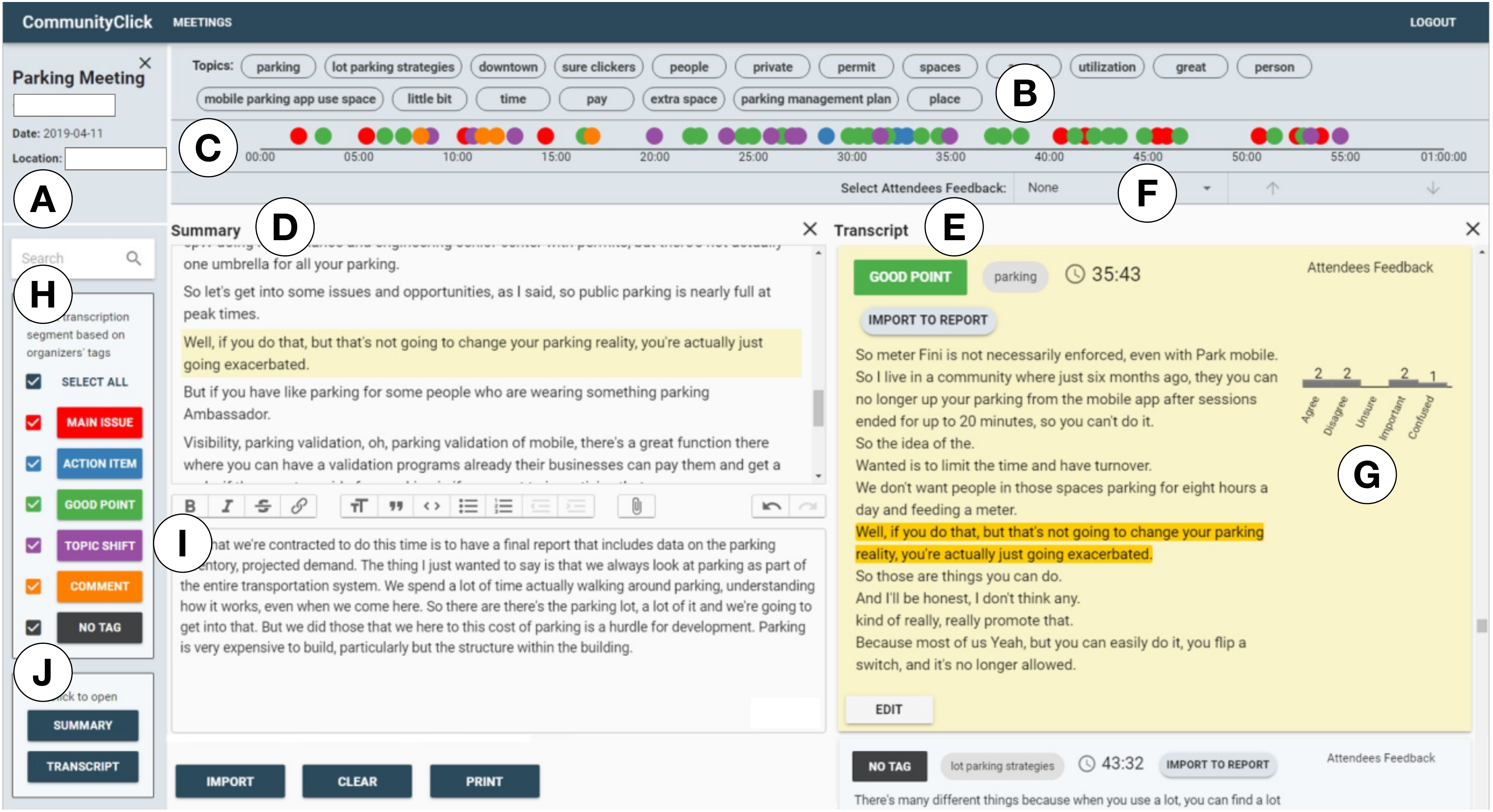}
\caption{A snapshot of CommunityClick's interface. A) The title provides useful metadata about the meeting such as date and location. B) The main topics extracted from the meeting transcript. C) The timeline visualizes organizers' tags in chronological order. Each circle represents a tag. Clicking on a circle brings organizers to the corresponding transcript segment. D) The interactive feedback-weighted summary. E) Transcript view displays the transcript text alongside organizers' assigned tag, the main topic, and aggregated attendees' feedback in that time interval for each segment. F) Filters for attendee's feedback (based on what was provided on iClicker). G) The bar chart displays attendees' feedback. H) Filters for organizer's tags. I) Rich text editor for organizers to author the report. J) Options to view or collapse the summary or the transcript view.}
\label{fig:interface}
\end{figure}

To transcribe the meeting audio, we used automatic speech recognition techniques from Assembly AI~\cite{assemblyai}. We assessed the quality of this method by comparing them to human-generated reference transcripts. We found our approach to be on-par with human-generated transcripts. The results of these analyses are presented in full in the \hyperref[appendix]{Appendix} section. We combined the transcript with the timestamped tags and feedback to transform the recorded meeting audio into timestamped text. Furthermore, we used the organizers' tags to divide the meeting transcript into manageable and consumable segments. Previous work showed that there is a gap (2 seconds on average) between hearing something, registering it, and taking actions upon it, such as clicking a button for annotation~\cite{risko2013collaborative}. Based on prior work and our early pilot experiments, for each organizer's tag, we created a 30 second time window around the tag (2 seconds before the tag and 28 seconds after the tag). The complete meeting transcript is divided into similar 30-second segments. For each segment, we collected the attendees' feedback provided within that time-window. Consequently, the meeting audio is transformed into timestamped segments, each containing transcribed conversation, organizers' tags, and a set of attendees' feedback (Fig.~\ref{fig:interface}(E)). We also extracted the main discussion points from the transcript segments using Topic Rank~\cite{bougouin2013topicrank} (Fig.~\ref{fig:interface}(B)). We chose this topic modeling method to have better multi-word topics  which are useful for better topic representation~\cite{blei2009visualizing}.

\subsubsection{Applying text summarization method to incorporate attendee's feedback into meeting discussion summaries (G3)}
To summarize the meeting transcript, we used the graph-based TextRank algorithm~\cite{mihalcea2004textrank}, which is based on a variation of the PageRank~\cite{page1999pagerank} algorithm. TextRank is known to deliver reasonable results in summarizing meeting transcripts~\cite{garg2009clusterrank} in unsupervised settings. However, it treats all input text the same without any domain-specific consideration. Our goal was to incorporate attendees' feedback in the summarization process so that the resultant summary was weighted by attendees' feedback. To that end, we added two critical modifications to the original methodology: 1) by incorporating attendees' feedback while computing relative importance of sentences, and 2) replacing the vanilla similarity function used in TextRank with a bag-of-words ranking function called BM25, which is proven to work well in information extraction tasks~\cite{robertson1995okapi}. 

Each individual transcript is treated as a set of sentences ($s_1, s_2...s_n$). Each sentence is considered as an independent node. We used a function to compute the similarity between these sentences to construct edges between them. The higher the similarity between the sentences, the more important the edge between them will be in the graph. The original TextRank algorithm considers the relation between two sentences based on the content (tokens) they share. This relationship is  between two sentences $S_i, S_j$, for every $w_k$ common token, is given by equation~\ref{tr_sim}.
\begin{equation}
    \centering
    sim(S_i, S_j) = \frac{\vert \{w_k \vert w_k \in S_i \& w_{k} \in S_j\}\vert}{\log(\vert S_j \vert) + \log(\vert S_i \vert)}
    \label{tr_sim}
\end{equation}
We replaced the above similarity function by a BM25 ranking function utility defined by equation \ref{bm_sim}.
\begin{equation}
    \centering
    sim(S_i, S_j) = \sum_{k=1}^n IDF(w_k \in S_i) \frac{f(w_k \in S_i, S_j). (a + 1)}{f(w_k \in S_i, S_j) + a.(1-b+b.\frac{\vert P \vert}{\mu_{DL}})}
    \label{bm_sim}
\end{equation}

\noindent where $a, b$ are function parameters ($a=1.2, b=0.75$), $f(w_k, S_j)$ is $w_k$'s term frequency in $S_j$, $IDF$ is the inverse document frequency, and $\mu_{DL}$ is the average length of the sentences in our collection. More importantly, since we timestamped and tagged our transcripts, we knew which instances were potentially more important in terms of garnering attendees' feedback. To incorporate those, we augmented every edge weight $w_{i,j}$ by a factor of $\epsilon$ determined experimentally (set to $1.10$ if either of the sentences being compared prompted feedback, or $0.90$ otherwise). From the constructed graph of sentences, we computed the final relative importance of every vertex (sentence), and selected the top-n sentences corresponding to about $30\%$ of the total length of the transcript which were then presented, in their original order, as the summary of the meeting discussion. 

We performed a series of ablation tests to evaluate the robustness of our summarization approach. The results of these tests are presented in full in the \hyperref[appendix]{Appendix} section. For three different meeting transcripts that we had generated using CommunityClick, we quantitatively evaluated the auto-generated summaries against human-annotated reference summaries using the widely used ROUGE~\cite{lin-2004-rouge} metrics. Across different meeting transcripts, we observed similar ROUGE scores, indicating that we produced summaries of consistent quality of summaries across different meetings. Additionally, we evaluated our algorithmic approach on the popular AMI meeting corpus~\cite{carletta2005ami}. Our results were found to be comparable to the current state-of-the-art methods~\cite{zhong2019searching, shang2018unsupervised} employed on the AMI dataset under unsupervised settings.

\subsubsection{Enabling exploration of an augmented meeting transcript through an interactive interface to help organizers author meeting reports (G4)}
We developed CommunityClick's interface as a web application. It allows multi-faceted exploration and analysis of meeting data by providing components including the title, filters, discussion topics, timeline, summary, text editor, and finally the augmented meeting transcript segments (Fig.~\ref{fig:interface}). The title contains the metadata about the meeting, including the meeting title, date, and location (Fig.~\ref{fig:interface}(A)). The filters allow organizers to explore the transcript segments according to the selected feedback or tags of interest (Fig.~\ref{fig:interface}(F, H)). These options are customizable, and organizers may customize the tags to suit their purpose before the meeting. We chose to visually collapse transcript segments that are filtered as opposed to completely removing them from the view to communicate to the users that there are additional conversations that transpired between the segments currently visible.

In the topic and timeline component, we provide the list of most relevant topics and the timeline of the meeting discussion (Fig.~\ref{fig:interface}(B, C)). The organizers can filter the transcript segments based on any topic. The timeline displays the organizers' tags using circles in a chronological manner, where each circle represents a tag, and the color corresponds to organizers' tags (Fig.~\ref{fig:interface}(C)). This provides the organizers with a temporal distribution of tags that demonstrates how the conversation progressed during the meeting. When a circle is selected, the transcript is scrolled to the corresponding segment and highlights the background to distinguish it from other segments.

The feedback-weighted extractive summary is presented in a textbox (Fig.~\ref{fig:interface}(D)). Each of these sentences is interactive, and upon selection, they navigate to the transcript segment it was extracted from. This can enable organizers to explore the transcript and get a better understanding of why the sentence was added to the summary. Below the summary, we added a rich text editor for authoring the meeting report with rich formatting options (Fig.~\ref{fig:interface}(I)). We also added options for attaching additional files or images. Once the report is created, it can be printed in PDF format directly, without switching to other external printing applications. 

Finally, we present the augmented transcript divided into transcript segments (Fig.~\ref{fig:interface}(E)). The segments are ordered chronologically. Each transcript segment contains the transcript text, associated organizer's tag, the most relevant extracted topic, time of the segment, option to import the summary of the selected transcript to the text editor, and aggregated attendees' feedback in the form of a bar chart. For easy tracking, we highlight the transcript text that are added to the summary. Organizers can edit the segments to assign or change tags and topics. However, they do not have control over attendees' feedback to mitigate bias injection. To reduce clutter on the screen, we added two additional filters to collapse the summary or augmented transcript (Fig.~\ref{fig:interface}(J)). 

\subsection{Pilot Study}
To explore whether CommunityClick could be effectively deployed in a real-world town hall meeting, we performed a pilot study where we simulated a town hall with nine participants. We recruited eight participant as meeting attendees and one participant as the organizer, who had previous experience with organizing meetings. We refer to the attendees who participated in our pilot study as \textbf{PA} We recruited all participants using word of mouth from a public university in the U.S. For the discussion topic, we selected two contentious issues regarding the university that were popular at the time of our pilot study. The topics included discussions around building a new common room for graduate students and the rise of racist activities across the campus. All participants were graduate students (6 males and 2 females with a average age of 27.25). The goal of the pilot study was to assess the system workflow for potential deployment and whether the attendees could share their feedback silently using iClickers without interrupting others. Furthermore, we used the augmented transcript from the meetings to enable the pilot study organizer we recruited to explore the meeting discussions to identify potential interface issues. 

The meeting took 60 minutes which is similar to a traditional town hall. We collected 292 items of feedback from attendees (avg 36.5 per attendee $\pm$ 8.23) and 56 tags from the organizer. After the meeting, we asked attendees to share their experiences of using iClickers to voice their opinions using open-ended questions. The findings suggested that attendees found iClickers easy to get used to. They were able to share their feedback silently using iClickers, managed to avoid potential confrontations, and thought they could contribute more compared to their experiences in other meetings. However, one attendee mentioned about difficulties around remembering which iClicker button mapped to which attendees' tag. Another attendee mentioned that while expressing strong agreement or disagreement with the ongoing discussion, some attendees might spam the iClicker button which might \blockquote{\emph{dilute the opinion values}} (PA-4). The organizer mentioned using iClickers enabled him to focus more on the discussion and the interface allowed him to better capture attendees' feedback. He recalled that some attendees were silent, but the bar charts showed feedback from all eight participants, meaning they were participating silently. He also identified the flow of the discussion, and important discussion points. 

This pilot study helped us to better understand and solidify operational procedures to perform real-world deployment of our system. Based on the feedback we received, we modified the system and user interface. For instance, we added a spamming prevention technique~\cite{sun2013synthetic} by calibrating the system to capture one click from each attendees' iClicker in a 30-second window to negate the possibility of diluting the values of specific feedback options. Furthermore, we added an option to collapse the summary or transcript to reduce interface clutter. Finally, we used written labels on iClickers to reduce attendees' cognitive load in remembering the mapping of response options. 

\section{Evaluation}
\label{evaluation}
We evaluated the application of iClickers as a real-time response system, particularly for the ability of silent attendees to share feedback, and the efficacy of our approach in enabling organizers to explore, capture, and incorporate attendees' feedback to author more comprehensive reports. To that end, we conducted a field experiment to examine if the attendees could effectively use iClickers to voice their feedback. In addition, we followed up by conducting semi-structured interviews with 8 expert organizers to evaluate if CommunityClick could enable them to capture attendees' feedback and generate more comprehensive reports.

\subsection{Field Experiment: Parking Town Hall}

We deployed CommunityClick at a town hall in a college town in the U.S. The meeting focused on a new set of proposals to improve the parking condition of the town. We reached out to the town officials a month before the meeting took place. They allowed us to deploy our system, record the discussion, and use the data. They also introduced us to the organizer who facilitated the meeting. We explained the procedure to them and discussed the tags to be used for both the organizer and attendees. The town hall took place on a Thursday evening and was open for all to attend. 

\subsubsection{Meeting Participants} 
There were 31 attendees and 1 organizer present in the meeting. We provided attendees' iClickers with Agree, Disagree, Unsure, Important, and Confused tags to 31 attendees. For the organizers, we provided them iClickers labeled with Main Issue, Concern, Supportive, New Idea, and Good Point tags as per our pre-meeting discussion. 

\subsubsection{Procedure}
At the beginning of the town hall, we provided a brief tutorial for five minutes on how to use the iClickers to the meeting attendees. We also received consent from the attendees about recording their discussions. The meeting began with an organizer presenting the meeting agenda and the new parking proposals to the meeting attendees. The attendees and the organizer used iClickers to tag the conversations throughout the meeting. After the presentation, the attendees engaged in discussing the proposals. The meeting lasted for 76 minutes. At the end of the meeting, we provided post-study questionnaires to attendees that asked various questions, such as their reasons behind attending the meeting, their usual experience during town halls, whether they could share their opinions by speaking up, and how did using iClickers compare to such experiences. They responded on a five-point Likert scale. We also asked them open-ended questions around their experience of working with iClickers, whether they faced any issues or challenges, and suggestions to improve their experiences and our approach. The post-study questionnaire is provided as a supplementary material. 

\subsubsection{Data Collection and Analysis}
We were given permission to collect and use the data from the town hall by government officials. We collected 61 minutes of meeting audio for transcription. We also collected organizer's tags and attendees' feedback from the meeting. In total, we captured 56 tags from the organizer. Out of 31 meeting attendees, 22 used the iClickers we provided to share their feedback. Out of these 22 attendees, 20 of them filled up the post-study questionnaire. We report the statistics based only on these 20 attendees' responses. We captured a total of 492 attendees' feedback with an average of 24.6 feedback items per attendee with a standard deviation of 6.44.  This data was later used to populate CommunityClick's interface for demonstrating its various functionalities to meeting organizers, which we describe in~\ref{sub:interview}. We also collected the post-study questionnaire responses and entered them into spreadsheets for creating charts (Fig.~\ref{fig:field_exp}) and statistics for analysis. 

\subsubsection{Findings}

From the analysis of the attendees' iClicker usage patterns, we found that the attendees' used the tag \textit{Agree} the most 187 clicks (38\%), followed by \textit{Important} with 103 clicks (21\%), \textit{Disagree} with 93 clicks (19\%), \textit{Confused} with 79 clicks (16\%), and finally \textit{Unsure} with the least amount of 30 clicks (6\%) only. Initially, we were surprise to see the large gap between agreement and disagreement. However, upon closer inspection, we found that on several occasions, the attendees who were using iClickers was clicking \textit{Agree} when other vocal attendee were verbally expressing their disagreement to a discussion topic. This behavior pattern indicates that the silent attendees used iClickers to provide their support for an ongoing argument alongside sharing their own opinions. We also found that the attendees did not press any iClicker options during the introduction when the organizers were setting up the discussion and conclusion of the meeting when the organizers expressed gratitude for attending the meeting and other social conversation. This suggests that the attendees took their opinion sharing using iClickers seriously and did not randomly clicked different options during the meeting. 

\begin{figure}
\centering
\includegraphics[width=1\textwidth]{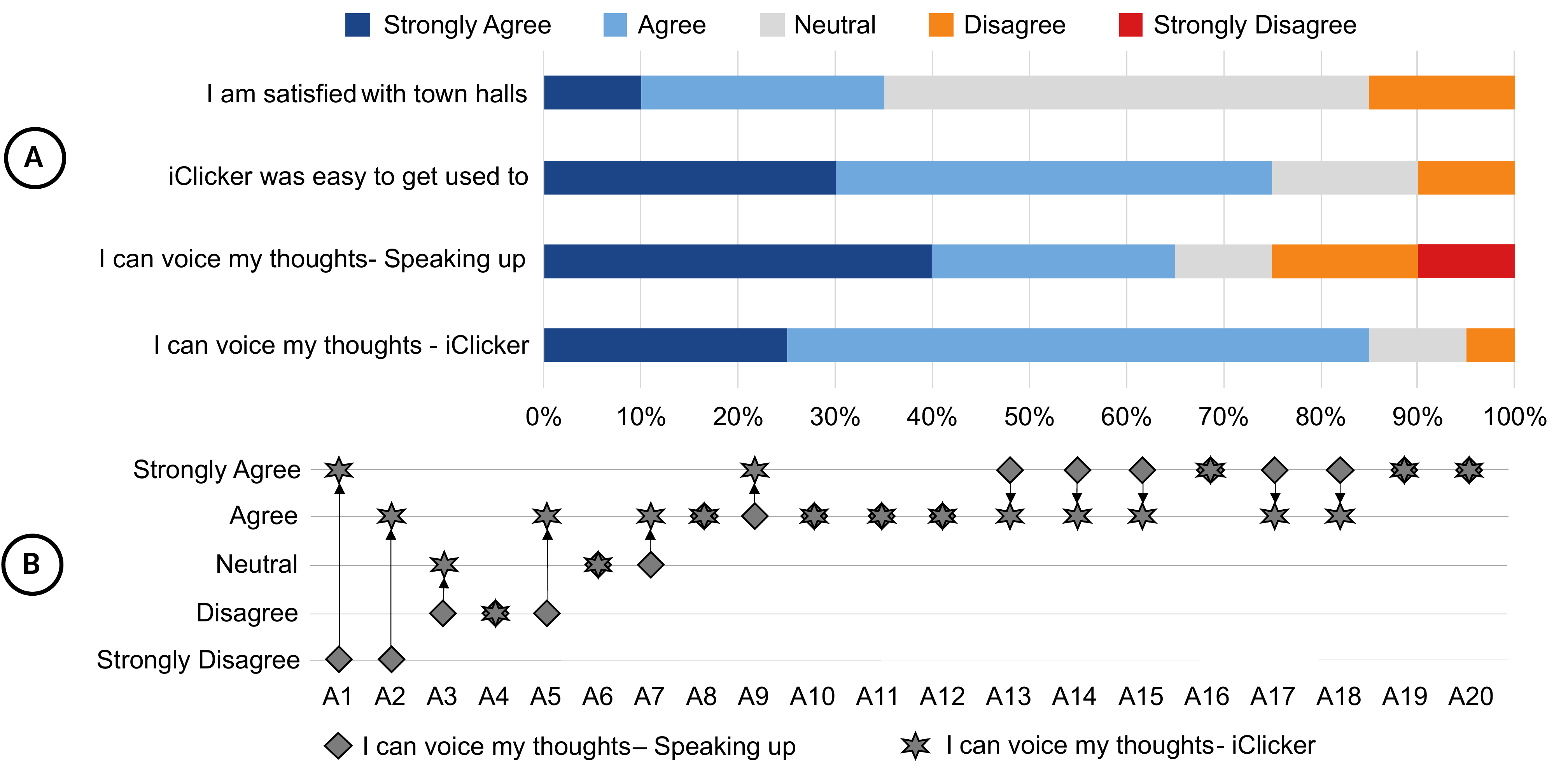}
\caption{The results from the field experiment. A) Attendees' responses show that the majority of meeting attendees were not satisfied with the status quo of town halls but found iClickers easy to get used to. It also displays the number of attendees who thought they could share their voices by speaking up or using iClickers. B) A deeper comparison between speaking up and using iClickers to share attendees' feedback. The diamonds (\textcolor{gray}{\ding{117}}) and stars (\textcolor{gray}{\ding{86}}) represent 20 attendees' (A1-A20) responses to questions that asked them to rate their experiences of sharing opinions by speaking up and using iClickers respectively during town halls. The arrows show the difference and increase or decrease in their ratings. The arrows demonstrate that the majority of the participants who were not satisfied with the current methods of sharing opinions (A1-A3, A5) during town halls found iClickers to be a better alternative. They also show the participants who were comfortable with speaking up during meetings, did not endorse iClickers as strongly to share their voices.}
\label{fig:field_exp}
\end{figure}

From the analysis of the post-study questionnaires, we found that all of the 20 attendees either lived or worked in the town of Amherst, where the town hall was organized. 95\% of these attendees (19 responses) were well-accustomed with such meetings and they mentioned attending similar town halls twice a year, while 50\%  (10 responses) attended these meetings more than five times per year. When asked about their exposure to technology, all meeting attendees responded to owning at least one computer and one smartphone with an internet connection, which they were comfortable using. However, their responses to the experiences in town halls varied as presented in Fig.~\ref{fig:field_exp}(A). 25\% of attendees (5 responses) mentioned that they did not feel that they are able to voice their thoughts in town halls. Only 35\% (7 responses) were pleased with the way such town halls are organized while 50\% of the attendees (10 responses) were neutral in their responses. 75\% attendees (15 responses) responded that they got used to iClickers quickly. 85\% mentioned (17 responses) they were able to share their thoughts using iClickers compared to only 65\% (13 responses) attendees who are comfortable with speaking up to share opinions. The majority of the attendees (90\%) were positive about their experiences of using iClickers to share their opinions (18 responses). One attendee (A9) mentioned, \blockquote{\emph{I feel like I could contribute more than usual and I would definitely like use it in future meetings.}} 

We further compared the attendees' responses on their ability to share their thoughts in town halls by voicing their opinions against using iClickers (Fig.~\ref{fig:field_exp}(B)). The data shows that almost all of the attendees who did not think they could share their thoughts by speaking up, except for one (A4) thought they could voice their opinions using iClickers. One such attendee (A2) mentioned, \blockquote{\textit{I didn't like what others were saying. But instead of interrupting, I just used the clicker to say that I didn't agree with them.}} We also found that, while agreeing that iClickers' could provide a way to voice opinions in town halls, the attendees who strongly preferred speaking up did not rate their experience of using iClickers to share opinions as high. One of these attendees (A17) mentioned, \blockquote{\textit{I was distracted, so I didn't use it that much.}} 

We identified two important insights from this field experiment (Fig.~\ref{fig:field_exp}). Firstly, it demonstrated that the silent attendees' who could not speak their minds found a way to voice their opinions without apprehension of confrontation in town halls using iClickers (Fig.~\ref{fig:field_exp}(A) and (B), attendees A1-A3, A5). However, we also found that some attendees who strongly agreed that they were satisfied with the current method of sharing opinions by speaking up, did not as strongly endorse iClickers as a way to share their opinions (Fig.~\ref{fig:field_exp}(B), attendees A13-A15, A17, A18). We speculate two reasons for such reduced ratings for iClickers. First, for the attendees who are already comfortable with speaking up, iClickers might seem like an additional step to share opinions which might lead to distractions, as mentioned by one of the attendees (A17). The second reason might be a reluctance to deviate from the norm and use technology in an established albeit impaired town hall proceedings and customs. Nevertheless, our results suggest that the addition of iClickers could be an acceptable trade-off between providing the silent attendees a way to communicate their opinions and mildly inconveniencing the adept and vocal meeting attendees.

\subsection{Semi-structured Interviews with Meeting Organizers}
\label{sub:interview}
We conducted semi-structured interviews with 8 expert meeting organizers, who were experienced with organizing or facilitating town halls to gather data on community's needs, issues, and ideas. They were also adept at compiling meeting reports that play a pivotal role in informing civic decision-making. Our objective was to examine if CommunityClick's interactive interface could help the organizers to better capture the attendees' feedback to author more comprehensive reports that preserve the equity and inclusivity of voiced opinions in town halls.

\subsubsection{Participants}
We reached out to a total of 29 expert organizers from across the U.S. 8 of them responded by agreeing to help us with evaluating CommunityClick. Our interviewees were experts in their fields with intimate knowledge of town hall organization and decision-making. We refer to our semi-structured interview participants as \textbf{P}. On average, they had over 20 years of experience. One interviewee (P1) was the organizer from our field experiment---the town hall on parking. We made connections with the others (P3-P8) during our formative study. All of our interviewees were based in the U.S.  

\subsubsection{Procedure}
Several experts we engaged with to evaluate CommunityClick were excited about its potential and agreed to deploy the system in their then upcoming town halls. Our original evaluation plan involved several deployments in the wild, then providing organizers with the meeting audio from these deployments and ask them to write two reports, one using their current method of writing reports, and the other using CommunityClick's augmented meeting transcript and interactive interface. We wanted to study the differences between these reports to investigate the efficacy of our system. However, the recent COVID-19 pandemic forced the organizers to cancel all town halls until further notice, and we were compelled to cut our evaluation short. Due to this setback, we revised and modified our evaluation procedure as follows.

We deployed the CommunityClick interface on a public domain and shared it with our interviewees via emails at least two weeks before our interviews. To maintain privacy of usage, we provided each organizer with their own user account and login credentials. We also provided detailed instructions on how to use CommunityClick's various features and encouraged the interviewees to explore the interface at their own convenience. We populated the interface with the data collected from the simulated meetings from our pilot study as well as the meeting from our field experiment for the interviewees to explore. During the interview sessions, we asked them open-ended questions focusing around their current practices towards town hall organization, how using CommunityClick differed from these practices, how useful could CommunityClick be to capture silent attendees' feedback and marginalized perspectives, could the interface allow them to author better reports, and finally, suggestions to improve CommunityClick. We also allowed them to ask any questions they might have about CommunityClick. The interview questions are provided as a supplementary material. We conducted the meetings over video conferencing via Zoom~\cite{zoom}. The interviews lasted between 45-60 minutes. All participation was voluntary. Each interview was conducted by an interviewer and a dedicated note-taker from our research team. 

\subsubsection{Data Collection and Analysis} We transcribed over 400 minutes of audio recording from our interviews with organizers. We also took extensive notes throughout the meeting. Finally, we thematically analyzed the interview transcripts and notes taken using the open-coding method~\cite{burnard1991method}. Two members of our research team independently coded the data at the sentence level using a spreadsheet application. The inter-coder reliability was 0.89 which was measured using Krippendorff's alpha~\cite{krippendorff2011computing}. We had several iterations of discussions among the research team to condense the codes into the themes presented in Table~\ref{tab:theme_table}.

\begin{table*}
    \caption{This table shows themes that emerged from analyzing the interviews with the organizers. The codes associated with the themes and their description is also presented in the table.}
    \scriptsize
    \setlength\tabcolsep{6pt}
    \ra{1.5}
    \centering
    \begin{tabular}[t]{p{2.6cm} p{6cm} p{4cm}}
    \toprule
    \textbf{Themes} & \textbf{Codes} & \textbf{Descriptions}\\
    \midrule
    Enabling inclusivity & Equitable platform, problem speaking up, opinion sharing, understanding others, inclusive opinions & CommunityClick's impact on inclusivity in town halls\\
    Diverse perspectives & Shared narrative, honest reflections, attendee's reactions, identifying conversation flow & Different perspectives and opinion shared in town halls\\
    Report quality & meeting summarization, missing information, credible process, comprehensiveness, accurate representation & CommunityClick's utility in creating reports\\
    Meeting organization & Unstructured discussions, real-time attendee's response, tracking response, customized tags, measuring consensus & Organizing meeting-generated data\\
    Interface learnability & Intuitiveness, easy-to-use, formatting, data exploration & Users' ability to learn and use interface features\\
    Concerns and caveats & Technology as a barrier, tech-savvy, distraction factors, young generation & Concerns regarding CommunityClick's usage in town halls\\
    Improvement suggestions & Real-time feedback, opinion statistics, organizers' input & Suggestions to improve our approach\\
    \bottomrule
    \end{tabular}
    \label{tab:theme_table}
\end{table*}

\subsubsection{Findings}
Our analysis of interview transcripts and notes surfaced critical insights on how CommunityClick could enable attendees to share opinions and help organizers to capture inclusive feedback to author more comprehensive reports. We elaborate on these insights in the following and discuss possible room for improvement.
\\\\
\noindent\textbf{CommunityClick can create a more inclusive platform to share opinions.} Our interviewees were unanimous (8 out of 8) in acknowledging CommunityClick's potential to create an inclusive platform for community to share their opinions. They shared with us several example scenarios they experienced where CommunityClick could have provided silent attendees a way to speak their minds. P1 mentioned, \blockquote{\textit{People want to share their opinions, but sometimes they just can't express themselves because they're not comfortable talking, or they're nervous about how they'll appear to or who is around the table. Often they are intimidated. Here, }"intimidated" \textit{is a strong word. But I don't think it's the wrong word.}} P2 mentioned, \blockquote{\textit{There was an individual who attended several meetings, it was clear that their presence had an impact on people's willingness to speak at all, or the opposite effect, where people escalated in reaction to that person. Giving them the ability to click help both ways. They can avoid confrontation or avoid escalation by just clicking.}} Similarly, P3 drew examples from his experiences, saying, \blockquote{\textit{Even if the attendees are from the U.S., [people with] different upbringings or cultural backgrounds have a disadvantage to those who are quite familiar with the moors of group dynamics. In our town halls, we only take votes on questions or get agreements, but in a conversation, there are so many euphemisms, colloquialisms, and metaphors that make it difficult for someone unfamiliar with them to understand others' reactions. There is real value in using options like ``confused'' and ``unsure'' to allow them to record that they didn't understand the conversation instead of forcing them to agree or disagree.}} P6 found further value in separating the attendees' tags and the organizers' tags to establish organizers' credibility. She mentioned, \blockquote{\textit{The organizers cannot unintentionally skew the attendees' feedback because [their tags] are separate. That way, we know the recorded feedback is unbiased.}} 
\\\\
\noindent\textbf{The augmented transcripts provide evidence of attendees' reflections.}
One of our primary goals was to enable organizers to have access to attendees' perspectives to form a shared narrative of the meeting discussions. After exploring CommunityClick's interface, the majority of interviewees (7 out of 8) mentioned how it enabled them to capture meeting attendees' reflections on the meeting agenda. P6 mentioned, \blockquote{\textit{It provides a way of ensuring that voices and reactions are reflected as people speak and click. It is a huge step towards having a more honest reflection of what really went on in the meeting.}} P3 further emphasized how CommunityClick not only captured the attendees' feedback but also allowed navigation of the conversation flow using the Timeline, \blockquote{\textit{This tool allows me to see both how many ideas have traction or agreement and how many don't, but just as importantly, how the flow went. The facilitators are concerned with the way topics are discussed in town halls. These topics are influenced by the surrounding conversations. It [Timeline] allows me to see reactions that might or might not be intended because of the sequence of conversations. Having a way to track that has a huge value.}} 

Regarding the interactive augmented transcript, P4 specifically preferred the way it enabled her to track attendees' responses. She drew a comparison with her usual methods for note-taking during town halls saying, \blockquote{\textit{We usually have a table facilitator and then a table observer. The table observer takes detailed notes, but it adds to the number of staff we have to have. So that creates an additional challenge, but the speech to text transcription makes a big difference in recording people's reactions. With [CommunityClick], maybe we won't need a table observer.}} P5 also mentioned how CommunityClick gave credence to attendees' reactions during the meeting discussions through the feedback bar charts. She said, \blockquote{\textit{It makes a lot of sense to see where people are aligned, where the challenges are, and giving information from their reactions. When changing policies, we hear from only a few voices who are either for or against something, and they tend to dominate the conversation. Having a visual and data-driven way to show what was the actual participation is gold. Sometimes people feel that a proposal is not aligned with their ideas. With the bar chart, you can show them that maybe you are the outlier and others agree with the proposal.}}
\vspace{0.3cm}
\noindent\textbf{CommunityClick can help create more comprehensive and accurate reports.}
All of our interviewees had prior experiences of writing reports by summarizing the meeting discussions and identifying key discussion points. They found various aspects of CommunityClick useful to not only author more comprehensive reports but also more accurate ones that lend credibility to the report creation process. P1 drew parallels with his experience of working in scenarios where designated note-takers took notes and his team generated the reports from those notes. He mentioned, \blockquote{\textit{People who take notes have varying abilities and the notes vary in quality. Instead, as you are writing reports, you have [CommunityClick], where you can see and add the reactions to what [attendees] discussed right away, it builds credibility for the process.}} P3 echoed similar sentiments, saying, \blockquote{\textit{You are usually distracted by the conversation while taking notes, which means you might miss every third word at a particular moment, which could be the difference between agreement and disagreement. Having it transcribed and summarized will remind a facilitator of some things that he or she may not have remembered or make it more accurate, because they may have remembered it differently. I love the fact that the [text analysis methods] can capture that objectively for us.}} P4 also emphasized the usefulness of importing a summary to the text editor. She mentioned, \blockquote{\textit{Having the text editor where you can start writing the report and pull in pieces from the transcript could be really helpful, because then as you read through the transcript and you're writing about some themes, you can pull characteristic quotes that would really help bring in more evidence for claims for those themes.}} Furthermore, we found that the report creation process can take a few hours to a few days depending on variables such as the way notes were taken, the length of the meeting, report creators' skills, etc. P7 highlighted the reduced workload and efficiency that CommunityClick could provide, saying, \blockquote{\textit{There is a physical component of getting into it, typing it up, theming, organizing, and editing which always takes longer than anticipated, I can see some of those issues can be fixed with this.}}
\\\\
\noindent\textbf{CommunityClick preserves the flow of meeting discussions by establishing an implicit structure.} The majority of our interviewees (6 out of 8) thought CommunityClick could be best utilized to organize unstructured meeting discussions. They emphasized that contrary to asking meeting attendees to respond to specific questions in town halls, CommunityClick allowed attendees to respond whenever they wanted, creating an implicit structure to the meeting while preserving the flow of discussion. One interviewee (P1) mentioned, \blockquote{\textit{[CommunityClick] would provide the biggest benefit in more unstructured kind of discussions. If you have a town hall, where people are less likely to speak up, [tags] would be helpful to understand their reactions and help with the theming.}} Another interviewee (P5) mentioned, \blockquote{\textit{It's hard to keep track of many ideas, but the visual organization of information helps to gauge reactions and figuring out if we reached consensus. But most importantly, it helps me to see if there are any social or racial injustice components into the proposals where there can be negative reactions.}} They also found the option to customize the attendees' feedback and organizers' tags useful to adapt to different meeting scenarios. One interviewee (P3) mentioned, \blockquote{\textit{Words may mean different things in different meetings. Having the ability to label and customize [tags] individually would be a way for different organizations to adjust to that. Sometimes we want to know [attendees'] hopes and concerns, but other times, we just want to know if they agree or disagree.}} However, P7 raised a concern about larger meetings, saying, \blockquote{\textit{I think [CommunityClick] will be useful for smaller meetings, but if there are hundreds of people, and everyone is speaking over each other, I'm not sure if you will be able to cope with that.}}
\\\\
\noindent\textbf{The simplicity and learnability of the interface affords intuitive exploration.}
From a usability standpoint, all of our interviewees (8 out of 8) found CommunityClick's interface to be simple and straightforward to work with. P3 extolled the interface saying, \blockquote{\textit{It's very intuitive, simple, easy to use, and navigate after the fact, edit, and update. All user interface features look well-thought-out to me considering the inner workings are extremely delicate and complicated.}} P4 valued the rich editing options of the text editor. She said, \blockquote{\textit{The automatic summaries can be used as a starting point of the report, as an initial cut, and then I can delete the things that might not be very useful and build up the report by adding more to it and formatting it. I can clearly see a lot of thought was put into designing the interface.}} P5 thought that the interactivity of the timeline was useful for navigating the augmented transcript. She mentioned, \blockquote{\textit{When I started clicking on the buttons on the circles at the top [timeline], it was very intuitive, like, it just automatically brings you to the places where that correlates with the statements, so you understand what it's connected to.}} P6 further emphasized CommunityClick's potential as a record-keeping and meeting exploration system, saying, \blockquote{\textit{Everything is linked together. So in that sense, it makes intuitive and logical sense when I'm looking at the data. It will be a total game-changer for policymakers and community organizers.}}
\\\\
\noindent\textbf{Concerns around technology in town halls.}
Although our interviewees praised CommunityClick, some of them raised a few important concerns. P8 mentioned how technology usage could be troublesome in town halls, saying, \blockquote{\textit{It feels like the technology itself could be seen as a barrier, because a lot of people might not feel quite as comfortable, clicking on things and reacting to.}} On a different note, P4 raised concerns about the sense of urgency such technology might impose on the meeting attendees. She said, \blockquote{\textit{[Attendees'] reactions, they are decisions that are being made in the spur of the moment as they're hearing information. And it's such a complex, sociological, and psychological response to information.}} Her concern was whether the urge to immediately respond to someone's perspectives could inhibit the ability to contemplate and deliver a measured response. Further concerns arose from P5, who mentioned how younger meeting attendees might have an advantage in the town halls if the technology is heavily used, saying, \blockquote{\textit{Younger generations tend to use technology so much more easily. And they turn to it so much more easily than older generations.}}
\\\\
\noindent\textbf{Possible room for improvements.}
We received some feedback from our interviewees on how to improve CommunityClick. Some of these suggestions focused on adding more real-time components to CommunityClick that can further augment the ongoing discussions. For example, P3 mentioned, \blockquote{\textit{Right now, [CommunityClick] helps me to understand attendees' reactions after the fact. I think if the facilitators could see them in real-time as the attendees are clicking away, they might be able to steer the conversation better to be more fruitful and fair.}} Other suggestions involved adding functionalities to add the organizers' own topics on top of the automatically extracted topics. In that regard, P4 mentioned, \blockquote{\textit{I guess it depends on who is using the system, but we look to dive a little bit more and would want to maybe customize the topics and themes}}. 

\section{Discussion}
\label{discussions}
From our field experiment and interviews with experts, we found that CommunityClick could be used to create an equitable platform to share more inclusive feedback in town halls. Compared to prior approaches to utilize technology in town halls~\cite{murphy2009promotion, boulianne2018citizen}, CommunityClick enabled attendees to utilize iClickers to silently and anonymously voice their opinions any time during the meeting without interrupting the discussion or the apprehension of being shut out by dominant personalities. We extended the use of audience response systems in prior works by adding five customizable options that go beyond binary agreement or disagreement and modifying iClickers as a real-time response system~\cite{boulianne2018citizen, bergstrom2009vote}. The experts we interviewed valued options such as \textit{confused}, or \textit{unsure} to identify if attendees were disengaged or did not understand the conversation without forcing them to agree or disagree~\cite{Karpowitz2005DisagreementDeliberation, sanders1997against}. The customizability of both organizers' tags and attendees' feedback further added to the flexibility of our approach that could potentially increase adaptability in meetings with diverse agendas in both civic and other domains. Moreover, the automation and augmentation of speech-to-text transcription, extraction of most relevant discussion topics, and feedback-weighted summarization of meeting discussion could potentially eliminate the need for separate note-takers. According to the organizers we interviewed, it could reduce the manpower requirement significantly compared to established approaches~\cite{lukensmeyer21, lukensmeyer2002taking}. 

From organizers' perspectives, CommunityClick could help create more comprehensive and accurate reports that could provide evidence of attendees' reflections. Prior work experimented with annotations during face-to-face meetings~\cite{kalnikaite2012markup, kelkar2010livetagging} for memory recall. In our work, we empowered organizers to go beyond recollection of events during meetings by integrating their own customized tags and enabled them to capture a more comprehensive picture of the meeting discussion. Furthermore, our interactive summary and attendees' feedback visualization provided a visual and data-driven way to highlight attendees' viewpoints, and outliers on critical points of discussions. This could enable organizers to receive a more clear reflection of what occurred in the meeting and author more accurate reports based on tangible evidence rather than incomprehensive interpretation~\cite{mahyar2019deluge}, that could further lend credibility to the report creation process. Prior work highlighted concerns about accuracy in computational approaches to analyze meeting data~\cite{mcgregor2017moretomeeting}. However, from our interviews with experts, we found that the accuracy and comprehensiveness of meeting reports often depends on meeting length, method of taking notes during the meetings, and note-takes' skills. We posit that synchronization of meeting data and addition of inclusive attendees' feedback into the summary and interface will enable organizers to author more accurate reports with the added benefit of reduced manpower requirement. Although during our interviews with meeting organizers, some of them highlighted that the augmented transcripts, discussion topics, and summaries could only be accessed after the meeting is finished, we argue that the latency is an acceptable tradeoff for increased comprehensiveness, credibility, and accuracy in generating reports. Furthermore, enabling access to the variety of meeting generated data could also help to reduce both organizers' and attendees' distraction during the meeting~\cite{gordon2011immersive, appleton2005gis} and allow them to engage with the ongoing discussion. In particular, the separation of attendees' feedback and organizers' tags along with the evidence of attendees' feedback in meeting reports could pave the way to instill trust between the decision-makers in the local government and the community members, which is considered to be a wicked problem in the civic domain~\cite{corbette2019trust, corbett2018problem, harding2015, mahyar2019deluge}. 

\subsection{Marginalization in Civic Participation and the Role of Technology}
Marginalization can be broadly defined as the exclusion of a population from mainstream social, economic, cultural, or political life~\cite{given2008sage}, which still stands as a barrier to inclusive participation in the civic domain~\cite{mahyar2019deluge, dickinson2019cavalry}. Researchers in HCI and CSCW have explored various communitysourcing approaches to include marginalized populations in community activities, proceedings, and designs~\cite{dickinson2019cavalry, erete2017empowered, walsh2019ai+, mahyar2018communitycrit, kim2016budgetmap}. In this work, we added to this body of work by designing technology that included silent voices in civic participation to increase the visibility of marginalized opinions. Our field experiment and interviews with experts demonstrated the efficacy of our approach to enable, capture, and include silent attendees' participation in town halls, regardless of their reasons behind keeping silent (social dynamics, fear of confrontation, cultural background, etc.). Our work also answers the call for more inclusive data analysis processes and practices to augment computational approaches~\cite{rhody2016dig, d2020data} by proposing a text summarization method that included and prioritized attendees' feedback when generating meeting discussion summaries. Such inclusive data analysis techniques could reflect the community's opinions, identify social and injustice, and support accountability in the outcome of civic engagements~\cite{erete2017empowered}. However, designing communitysourcing technologies to include marginalized opinions and amplify participation alone may not be enough to solve inequality of sharing opinions in the civic domain~\cite{torres2007citizen, bovaird2007beyond}. Despite the success of previous works~\cite{erete2017empowered, lukensmeyer21, boulianne2018citizen}, technology is rarely integrated with existing manual practices and follow-ups of engagements between government officials and community members are seldom propagated to the community. This lack of communication might lead to the uncertainty of attendees on whether actions will be taken to reflect their opinions. As a result, the power dynamics between the government officials and the community remain unbalanced, especially for the marginalized populations~\cite{dickinson2019cavalry, erete2017empowered, dickinson2018inclusion, taylor2012empowering}. One way to establish the practicality of using technology to include marginalized opinions is to integrate them into existing processes to convince the officials of its efficacy. Our work provides first steps towards integration of marginalized perspectives, however, long-term studies are required to assess the possibility and feasibility of integrating public perspectives into the actual civic decisions. 

\subsection{Integrating CommunityClick in Today's Town Halls}
Our formative study with 66 attendees and field experiment with 20 attendees bore a striking resemblance regarding the attendees' ability to share their opinions in town halls. We found that 17\% (11 out of 66) of attendees from the formative study and 25\% (5 out of 20) of attendees from the field experiment were not comfortable to speak up in town halls and needed a way to share their opinions. However, similar to previous works~\cite{lukensmeyer21, gastil2008political}, some of the meeting organizers we interviewed were apprehensive towards depending solely on technology due to concerns around logistics involved with the procurement and maintenance of electronic devices such as iClickers, reliability concerns of using technology that require proper management, and unfair advantage towards newer generation who are more receptive to novel technologies. We argue that renewed motivation, careful design choices, and detailed instructions could help overcome the novelty barrier of technology even for people from older generations~\cite{vaportzis2017older, d2014three}. Based on our experiences from this study, we advocate for integrating technology with current face-to-face methods. We do not suggest complete replacement of traditional methods with technology, rather we suggest augmenting them with technology to address some of their limitations.

CommunityClick could be gradually integrated as a fail-safe or an auxiliary method to complement the current process. All the functionality of CommunityClick would remain operational while the organizers could take personal notes in parallel alongside their iClicker interactions. This would allow organizers to retain their current practices while taking advantage of augmented meeting transcripts, discussion topics, and summaries from CommunityClick to better capture and understand attendees' perspectives when compiling reports. Another way to integrate CommunityClick into current processes would be to provide statistics of attendees' feedback so that the organizers could track the discussion dynamics and facilitate the conversation accordingly to keep attendees engaged. However, prior works suggested that real-time visualization or displays can add to attendees' distraction, causing them to disengage from the ongoing discussion~\cite{gordon2011immersive, appleton2005gis}. To circumvent this issue, optional companion mobile applications could be introduced only for organizers to receive real-time feedback on attendees' iClicker responses so that they can make course corrections accordingly without distracting the attendees or influencing their opinions.

\subsection{Expanding CommunityClick to other Domains}
From our field experiment and interviews with the experts, we demonstrated CommunityClick's potential in elevating traditional town halls. We argue that our proposed data pipeline can be expanded to other domains where opinion sharing is important for successful operation. For example, in education, whether in classroom settings or in massive open online courses~\cite{kizilcec2013deconstructing, d2007mind, jones2013classroom}, it could enable students to share their feedback anytime without interrupting the class on whether they could understand a certain concept, or if they were confused and needed more elaboration, especially when the class size is large. For educators, it could allow them to track the effectiveness of their content delivery, class progress, and student motivation, which might help them to adjust the course curriculum more effectively and efficiently, instead of receiving feedback from students once every semester. More importantly, familiarity with iClickers in education eliminates the entry point barrier for technology making the system readily adoptable~\cite{whitehead2010usingiClicker, addison2009usingiclicker}. 

Similarly, CommunityClick could be used as a possible alternative to expensive commercial applications in meeting domains within the business and other corporate organizations~\cite{meetingKing, iCompassTech}. Similar to town halls, in a corporate setting, CommunityClick could provide text summary and direct evidence of attendees' feedback from the meeting transcripts for better decision-making. Automatic text summarization remains a challenging problem~\cite{tas2007survey}, especially when it comes to automatically summarize meeting discussions~\cite{gillick2009global}. Our summarization approach emphasized the importance of meeting attendees' feedback instead of purely text-based summarization approach that treated all sorts of text documents similarly~\cite{tas2007survey, mihalcea2004textrank}. We argue that this \textit{feedback-weighted} summary could be valuable in generating domain-specific contextual text summarization in other meeting genres. Furthermore, there are potential applications of CommunityClick as a record-keeping tool which might be particularly applicable in journalism where the journalists can utilize the iClickers to annotate the interview conversation with important tags and later review the augmented interview conversation using CommunityClick's interface to better organize and write news articles or interview reports. 

\section{Limitations and Future Work} 
Our evaluations suggested the efficacy of CommunityClick in providing a voice to reticent participants and enabling organizers to better capture the community's perspectives. However, we had only one real-world deployment of CommunnityClick at a town hall due to the pandemic. We will continue to engage with meeting organizers and deploy CommunityClick in town halls to study the long-term impact of CommunityClick and attempt to gradually integrate our approach in the predominantly manual town hall ecosystem. Also, we found that iClickers might be distracting as a new technology for some attendees in town halls as the findings from our field experiment suggested. Furthermore, the logistical issues associated with both hardware procurement and the unavailability of software APIs might be a hindrance to some communities. To circumvent these issues, low-cost alternatives to iClickers or fully software-based substitute audience response system applications could be utilized~\cite{voxvote, slido}. A fully software based solution could also enable attendees to provide open-ended textual feedback which CommunityClick does not allow in its current state. However, further comparative studies are required to assess the cost, efficacy, and applicability of such alternatives to replace iClickers that would provide the same benefit without incurring additional financial, computational, or cognitive overhead.  

There are several avenues to improve CommunityClick in the future. For example, it could be augmented with more real-time components including a companion application to deliver dynamic feedback statistics for organizers to access and utilize during the meeting. We will also explore novel methodologies to speed up the automatic speech to text transcription process to further reduce the time required for data processing. One approach could be to utilize parallel pipeline processing~\cite{chaiken2008scope} where audio signals from the meeting will be processed concurrently, which might reduce processing time significantly. 

To further provide evidence from attendees' feedback to help organizers when authoring reports, the audio of discussions could be added and synchronized with the transcript segments. It could enable the organizers to identify vocal cues and impressions from the attendees who spoke during the town hall to further contextualize the discussion segment ~\cite{mehrabian2017nonverbal}. In addition, we will investigate the possibility of tracking individual iClickers IDs without risking the privacy of attendees to anonymously identify potentially contentious individuals who might be pushing specific agendas or marginalize minority viewpoints in a discussion. However, further studies are required to understand the potential computational challenges that might arise with such extensions. 

To further improve the report generation, we will explore novel technologies in natural language generation~\cite{wen2015semantically} to automatically write meeting reports that could further reduce meeting organizers' workload. In addition to exporting the created reports, we will further enable exporting various statistics around attendees' feedback, organizers' tags, discussion topics, etc. to be added to the report or used separately in presenting the outcome of the town halls to decision-makers. 

\section{Conclusion}
In this study, we investigated the practices and issues around the inequality of opportunity in providing feedback in town halls, especially for reticent participants. To inform our work, we attended several town halls and surveyed 66 attendees and 20 organizers. Based on our findings, we designed and developed CommunityClick, where we modified iClickers as a real-time response mechanism to give voice to silent meeting attendees and reflect on their feedback by augmenting automatically generated meeting transcripts with organizers' tags and attendees' feedback. We proposed a novel feedback-weighted text summarization method along with extracting the most relevant discussion topics to better capture community's perspectives. We also designed an interactive interface to enable multi-faceted exploration of the summary, main discussion topics, and augmented meeting-generated data to enable organizers to author more inclusive reports. We deployed CommunityClick in-the-wild to conduct a field experiment and interviewed 8 expert organizers to evaluate our system. Our evaluation demonstrated CommunityClick's efficacy in creating a more inclusive communication channel to capture attendees' opinions from town halls and provide evidence of attendees' feedback that could help organizers to author more comprehensive and accurate reports to inform critical civic decision-making. We discussed how CommunityClick could be integrated into the current town hall ecosystem and possibly expanded to other domains.

\bibliographystyle{ACM-Reference-Format}
\bibliography{sample-acmsmall}

\clearpage

\appendix

\section{Appendix}
\label{appendix}
\subsection{Summary Evaluation}
To summarize augmented transcripts, we wanted a robust unsupervised summarization algorithm. To that end, we improved TextRank~\cite{mihalcea2004textrank} to create an attendee-feedback-weighted summarization algorithm. To test the effectiveness of this extractive summarization algorithm, we took three different transcripts obtained from the following meetings\footnote{All of which are presently analyzed in the CommunityClick interface.}:
\begin{enumerate}
    \item A meeting discussion about creating a common room for graduate students in the department.
    \item A meeting organized by public school officials discussing challenges to building renovations.
    \item A meeting held on a college campus discussing the rise in racist activities. 
\end{enumerate}

We evaluated these transcriptions with 4 different human annotators, who were specifically tasked with identifying sentences that best summarized the discussion transcripts. The annotators were graduate students working in the area of natural language processing. They were familiar with the literature and practices of text summarization tasks. We treated these human-annotated summaries as references and against each of those reference summaries, we evaluated a ROUGE~\cite{lin-2004-rouge} score metric for the auto-generated summary. ROUGE is a widely used metric for evaluating computer-generated summaries. It measures how much of the reference (human-generated) summary is our algorithm ``capturing''. We ideally wanted to strike a good balance between conciseness (precision) and the amount of information captured (recall) by our auto-generated summary and therefore, we computed the ROUGE F1-scores which in turn measure both, precision and recall. We evaluated ROUGE-1 F1 and ROUGE-2 F1 scores where ROUGE-N denotes N-gram comparisons of auto-generated and reference summaries. For example, ROUGE-1 refers to the overlap of unigrams between our auto-generated summary and the reference summary. Similarly, ROUGE-2 would denote the bigram comparisons. Upon evaluating our auto-generated summaries for the three selected transcripts against their respective reference summaries, we noticed nearly-consistency ROUGE scores. Our full results are summarized in Table \ref{tab:rouge}. These results suggest that our algorithm produced summaries of consistent quality across different meeting transcripts. 

\begin{table}[h]
\caption{\texttt{ROUGE-1} and \texttt{ROUGE-2} F1 scores of generated summaries ($T_n$) against reference summaries produced by human annotators ($A_n$).}
\setlength\tabcolsep{6pt}
\begin{tabular}{@{}lccccllll@{}}
\toprule
{\textbf{Transcriptions}} & \multicolumn{4}{c}{\textbf{Rouge-1 F1}} & \multicolumn{4}{c}{\textbf{Rouge-2 F1}} \\ \cmidrule(l){1-9} 
   & $A_1$   & $A_2$   & $A_3$   & $A_4$   & $A_1$   & $A_2$   & $A_3$   & $A_4$   \\ \cmidrule(r){2-5} \cmidrule(l){6-9}
$T_1$ (Common room discussion) & 51.29 & 45.83 & 48.12 & 42.41 & 34.56 & 33.23 & 24.64 & 30.98 \\
$T_2$ (School building renovations) & 48.23 & 12.34 & 44.43 & 49.19 & 28.21 & 31.39 & 29.79 & 23.98 \\
$T_3$ (Activities on college campus) & 68.21 & 70.61 & 52.28 & 55.09 & 33.94 & 35.90 & 27.29 & 21.93 \\ \bottomrule
\end{tabular}
\label{tab:rouge}
\end{table}

Our summarization method allowed us to generate short unsupervised summaries of meeting transcripts by modeling those transcripts as a graph with sentences being the nodes. A similarity function is used to build edges between those nodes. BM25~\cite{robertson1995okapi} is a ranking utility function widely used to measure similarity in state-of-the-art methods for information extraction tasks. To measure the effectiveness of using BM25 as a similarity measure in our case and to further assess the robustness of our summarization system on a widely used standardized dataset, we evaluated the effect of changing the similarity metric used in TextRank through some ablation studies. We carried out these experiments on the standard AMI Meeting Corpus\footnote{\url{http://groups.inf.ed.ac.uk/ami/corpus/}}, which is a widely used meeting-summarization benchmark. We used the traditional test set of 20 meetings, each associated with a human-annotated summary of approximately 290 tokens. We compared the vanilla similarity metric~\cite{mihalcea2004textrank} against the BM25 similarity metric and evaluated the ROUGE scores for the AMI test-set reference summaries. Results for these experiments are summarized in Table \ref{tab:sim_rougue}. We observe that our approach resulted in a 3-point gain in recall (amount of information captured) and a 2-point gain in precision over the vanilla similarity metric used in the TextRank algorithm on the AMI-meeting corpus.

\begin{table}[h]
\caption{Macro averaged results for different similarity functions within TextRank.}
\setlength\tabcolsep{6pt}
\begin{tabular}{@{}lccc@{}}
\toprule
\multicolumn{1}{l}{\textbf{Similarity Function ($f$)}} & \multicolumn{3}{l}{\textbf{AMI ROUGE-1}} \\ \midrule
                                                     & \textbf{P} & \textbf{R} & \textbf{F-1} \\ \cmidrule{2-4}
TextRank (with BM25)                                                 & \textbf{42.98}       & \textbf{35.75}       & \textbf{37.39}         \\
TextRank (with cosine-similarity)                                             & 41.21       & 32.66       & 36.54         \\
TextRank (vanilla)                                            & 40.11       & 33.48       & 36.27         \\ \bottomrule
\end{tabular}
\label{tab:sim_rougue}
\end{table}

As we evaluated the performance on a domain-specific dataset (AMI meeting corpus), we argue that these gains are consistent across our data as well. Furthermore, we found that our results were in-line with recent state-of-the-art methods~\cite{zhong2019searching, shang2018unsupervised} on unsupervised extractive summarization. 

\subsection{Automatic Speech Recognition Evaluation}
To assess how accurately our Automated Speech Recognition captures the meeting audio to generate meeting transcripts, we carried out a tests to evaluate the Bilingual Evaluation Understudy Score (BLEU)~\cite{papineni-etal-2002-bleu} on the same transcripts described in the previous section. A BLEU score is a metric for evaluating a generated sentence to a reference sentence and is often used in machine translation tasks. We adopted it to evaluate our transcripts by counting matching n-grams in the generated text to matching n-grams in the reference text. A score 1.0 indicates a perfect match and a lowest score of 0.0 would indicate a complete mismatch. Our reference transcripts were generated by a single human annotator who manually transcribed the meeting audio. We consistently observed high BLEU scores, indicating that tokens appearing in the reference transcript are also present in the generated transcript, thereby implying that our ASR system was indeed reliable. Our full results are summarized in Table \ref{bleu}. 

\begin{table}[h]
\caption{Evaluation of generated transcripts against reference transcripts that were manually generated by humans. BLEU-n scores here effectively indicate n-gram overlaps between sentences from reference transcript and generated transcript. }
\begin{tabular}{@{}lccc@{}}
\toprule
                  & \textbf{$T_1$ (16 mins)} & \textbf{$T_2$ (55 mins)} & \textbf{$T_3$ (75 mins)} \\ \midrule
\textbf{BLEU - 1} & 0.9028                & 0.9024                & 0.9046                \\
\textbf{BLEU - 2} & 0.8811                & 0.8753                & 0.8733                \\
\textbf{BLEU - 3} & 0.8152                & 0.8232                & 0.8381                \\
\textbf{BLEU - 4} & 0.7812                & 0.7910                & 0.7604                \\ \bottomrule
\end{tabular}
\label{bleu}
\end{table}

\end{document}